\documentclass[a4paper,11pt]{article}
\pdfoutput=1 

\usepackage{jheppub} 

\usepackage{pdfsync}
\usepackage[T1]{fontenc} 
\usepackage{graphicx}
\usepackage{amsmath}
\usepackage{amssymb}
\usepackage[dvipsnames]{xcolor}
\usepackage{centernot}

\def\U1mt{U(1)_{L_\mu-L_\tau}}
\def\bsmm{b \to s \mu \mu}

\def\ol{\bar}
\def\nl{\nonumber\\}

\title{\boldmath Scalar dark matter behind $b \to s \mu \mu$ anomaly}

\author{Seungwon Baek}
\emailAdd{sbaek@korea.ac.kr}
\affiliation{Department of Physics, Korea University, Seoul 02841, Korea}

\abstract{We construct a scalar dark matter model with $U(1)_{L_\mu-L_\tau}$ symmetry in which the dark matter
interacts with the quark flavours, allowing
  lepton non-universal $b \to s \ell \bar{\ell}$ decays. The model can solve  $b \to s \mu \mu$ ($R_{K^{(*)}}$) anomaly and
accommodate the relic abundance of dark matter simultaneously while satisfying the constraints from 
other low energy flavour experiments and  direct detection
experiments of dark matter. The new fields include vector-like heavy quarks $U$ and $D$, $U(1)_{L_\mu-L_\tau}$ breaking scalar $S$, as
well as the dark matter candidate $X_I$ and its heavy partner $X_R$. To explain both $b \to s \mu \mu$ anomaly and the dark
matter, {\it i)} large mass difference between $X_R$ and $X_I$ is required,  {\it ii)} electroweak scale dark matter and heavy quarks are favoured,
{\it iii)} not only electroweak scale but ${\cal O}(10)$ TeV dark gauge boson $Z'$ and $X_R$ are allowed.
}

\begin{document} 
\maketitle
\flushbottom

\section{Introduction}
\label{sec:intro}

The flavour changing neutral current (FCNC) processes are known to be sensitive to new physics (NP) because they first occur at
loop level in the standard model (SM) and therefore are sensitive to heavy physics in the loop.
The NP scale they can probe is usually much higher than the scale the LHC can
produce. And these indirect searches for NP are complementary to the collider searches. Among many FCNC processes, the $b \to s \ell\ell$
transition has been drawing much interest for the last several years
because of anomalies in $B \to K^{(*)} \mu \mu$ and $B_s \to \phi \mu\mu$ decays.

In particular SM predictions on the ratio of branching fractions
\begin{equation}
  R_{K^{(*)}}=\frac{{\cal B}(B\to K^{(*)} \mu^+ \mu^-)}{{\cal B}(B\to K^{(*)} e^+ e^-)},
  \label{eq:RK}
\end{equation}
are close to unity, signifying the lepton flavor universality (LFU) in the SM.
However, the measurements at the LHCb for $K$~\cite{Aaij:2014ora} and $K^*$~\cite{Aaij:2017vbb} are lower than unity at $2.3-2.6\sigma$ level.
Because the ratio (\ref{eq:RK}) is free from hadronic uncertainty, it would be a clear sign for NP, if this violation of LFU persists in future experiments.
Including other observables, such as an angular observable in $B\to K^* \mu^+ \mu^-$ and branching fraction
of $B_s \to \phi \mu^+\mu^-$ and $\Lambda_b \to \Lambda \mu^+\mu^-$,  the deviations from the SM predictions increase as large as about 
5$\sigma$~\cite{Altmannshofer:2017fio, Capdevila:2017bsm, Ciuchini:2017mik,Alok:2017sui}, which we will call $\bsmm$ anomaly.
At $m_b$ scale the $b \to s \ell \ell$ transition is described by the effective weak Hamiltonian 
\begin{align}
{\cal H}_{\rm eff} &= -{4 G_F \over \sqrt{2}} V_{ts}^* V_{tb} \sum_i (C_i^\ell O_i^\ell +
                     C_i^{\prime \ell} O_i^{\prime \ell})+h.c.,
\end{align}
where the relevant effective operators are 
\begin{align}
O_{7\gamma}^{(\prime)} &= {e \over 16 \pi^2} m_b (\bar{s} \sigma^{\mu\nu} P_{R(L)} b) F_{\mu\nu}, \quad
O_{8g}^{(\prime)} = {g_s \over 16 \pi^2} m_b (\bar{s} \sigma^{\mu\nu}  T^a P_{R(L)} b) G^a_{\mu\nu}, \nl
O_9^{(\prime)\ell} &= {e^2 \over 16 \pi^2} (\bar{s} \gamma_\mu P_{L(R)} b)(\bar{\ell} \gamma^\mu \ell), \quad
O^{(\prime)\ell}_{10} = {e^2 \over 16 \pi^2} (\bar{s} \gamma_\mu P_{L(R)} b)(\bar{\ell} \gamma^\mu \gamma_5 \ell). 
\label{eq:effective}
\end{align}
In the SM the un-primed operators dominate the chirality-flipped primed ones.
In the SM we obtain $C_{7\gamma}^{\rm SM} \simeq -0.294$, $C_9^{\rm SM}\simeq 4.20$, $C_{10}^{\rm SM}\simeq -4.01$ at $m_b$ 
scale~\cite{Bobeth:1999mk,Mahmoudi:2018qsk}. 
The results from global fitting analyses~\cite{Altmannshofer:2017fio, Capdevila:2017bsm, Ciuchini:2017mik,Alok:2017sui} show that sizable NP contributions to
$C_{9}^\mu$ and/or $C_{10}^\mu$ 
can explain the $\bsmm$ anomaly. 

In this paper we consider a NP model with $C_{10}^{\mu,{\rm NP}}=0$, in which case the best fit value for $C_9^{\mu,{\rm NP}}$  
is~\cite{Capdevila:2017bsm}
\begin{align}
C_9^{\mu,{\rm NP}}&=-1.11 \pm 0.17, 
\label{eq:C9_fit}
\end{align}
with a SM pull of $5.8 \sigma$.
In addition to the SM gauge groups we introduce a new gauge symmetry $\U1mt$ under which the 2nd (3rd) generation leptons are charged with $+1 (-1)$.
It is known that the theory is anomaly-free even without extending the SM fermion contents.
{Since the $\U1mt$ gauge boson $Z'$ couples to muon, it can make a contribution to the anomalous magnetic moment of muon
  $(g-2)_\mu$~\cite{Baek:2001kca,Banerjee:2018eaf}. But the $Z'$ should be very light ($\lesssim 400$ MeV) to fully accommodate the 
discrepancy between the experiments and the SM predictions in the $(g-2)_\mu$~\cite{Altmannshofer:2014pba}.
}
The model can also be extended to accommodate neutrino data~\cite{Baek:2015mna,Baek:2015fea,Singirala:2018mio,Asai:2018ocx}.
In Ref.~\cite{Baek:2017sew} we introduced a fermion dark matter (DM) model whose stability is originated from $\U1mt$ 
symmetry~\cite{Baek:2008nz}.
The model can also explain $\bsmm$ anomaly by introducing $SU(2)_L$-doublet scalar field,
and we showed that there is a strong interplay between the DM and $B$-physics 
phenomenology~\cite{Crivellin:2015mga,Belanger:2015nma,Allanach:2015gkd,Ko:2017quv,Ko:2017yrd,Ko:2017lzd,
Arnan:2016cpy,Altmannshofer:2016jzy,Kawamura:2017ecz,Assad:2017iib,Baek:2018aru,Darme:2018hqg,Barman:2018jhz,Rocha-Moran:2018jzu,Faisel:2018bvs,Vicente:2018frk}.
In this paper we consider a ``spin-flipped'' version of the model in Ref.~\cite{Baek:2017sew}.
We introduce two complex scalar fields $S$ and $X$: $S$ breaks the $\U1mt$ symmetry spontaneously by developing vacuum expectation
value (VEV) $\langle S\rangle$, while the lighter component $X_I$ is stable by the remnant discrete $Z_2$ symmetry after $\U1mt$ symmetry is broken spontaneously
and become a DM candidate.
To explain the $\bsmm$ anomaly as well in this model, we introduce a vector-like quark $Q$ which can mediate quark couplings to $Z'$ boson.
We will study the solution of the $\bsmm$ anomaly and the DM phenomenology in this model.

This paper is organized as follows. In Section~\ref{sec:model}, we introduce the model and calculate the new particle mass spectra. 
In Section~\ref{sec:NP} we calculate NP contribution to $b \to s \mu\mu$, and consider low energy constraints including 
$C_9^{\ell,  {\rm NP}}$,  $\Delta m_s$ in $B_s - \ol{B}_s$ mixing, $B \to K^{(*)} \nu \ol{\nu}$, $b \to s \gamma$, the anomalous magnetic moment of
muon $a_\mu$, and the loop-induced effective $Z b \bar{b}$ coupling. In Section~\ref{sec:DM} we consider dark matter phenomenology.
Finally we conclude in Section~\ref{sec:concl}. Loop functions are collected in Appendix~\ref{app:loop}.

\section{The model}
\label{sec:model}

We introduce a scalar dark matter candidate $X$ and a scalar boson $S$ which gives a mass to $U(1)_{L_\mu-L_\tau}$ gauge boson $Z'$ after the symmetry is
broken down spontaneously by the VEV of $S$. To couple the $Z'$ gauge boson to the quarks we also introduce a vector-like $SU(2)_L$-doublet
fermion $Q \equiv (U, D)^T$.
Their charges under the $\U1mt$  as well as those under the SM gauge groups  are shown in Table~\ref{tab:particles}.
\begin{table}[t]
\begin{center}
\begin{tabular}{|c|c|c|c|}\hline
\multicolumn{1}{|c|}{} & \multicolumn{1}{|c|}{New fermion} & \multicolumn{2}{|c|}{New scalars} \\ \hline\hline
                & $Q$             & $X$   &    $S$ \\ \hline
$SU(3)_C$  &  {\bf 3}        & {\bf 1}    &  {\bf 1}   \\ \hline
 $SU(2)_L$     &  {\bf 2}          & {\bf 1}  &  {\bf 1}   \\ \hline
 $U(1)_Y$ &  ${1 \over 6}$      & $0$      &  $0$   \\ \hline
 $U(1)_X$ & $q_Q(\equiv -q_X)$  & $q_X$ & $q_S (\equiv -2 q_X)$     \\\hline
\end{tabular}
\caption{New particles in the model with their quantum numbers under the gauge group $SU(3)_C \times SU(2)_L \times U(1)_Y \times \U1mt$. }
\label{tab:particles}
\end{center}
\end{table}
The Lagrangian respecting the gauge symmetry and charge assignments in Table~\ref{tab:particles} is written as
\begin{align}
  {\cal L} &= {\cal L}_{\rm SM} -V -{1 \over 4} Z'_{\mu\nu} Z^{\prime\mu\nu}
             -{\sin\chi \over 2} Z'_{\mu\nu} B^{\mu\nu} + \ol{Q} (i \centernot D -M_Q) Q
             + (D_\mu X^\dagger) (D^\mu X)  + (D_\mu S^\dagger) (D^\mu S) \nl
  &- \sum_{i=1}^3 (\lambda_i \ol{q}_L^i Q X + {h.c.}),
\label{eq:model}
\end{align}
where $D_\mu$ is the covariant derivative, $i(=1,\cdots,3)$ is the quark-generation index, and $A_{\mu\nu} =\partial_\mu A_\nu-\partial_\nu A_\mu$ ($A=Z',B$)
is the field strength tensor. The scalar potential is in the form,
\begin{align}
  V &= -\mu_H^2 H^\dagger H -\mu_S^2 S^\dagger S + m_X^2 X^\dagger X   +(\mu X^2 S + h.c.)+\lambda_H (H^\dagger H)^2  +\lambda_S (S^\dagger S)^2
      +\lambda_X (X^\dagger X)^2 \nl
      &+ \lambda_{HS} H^\dagger H S^\dagger S + \lambda_{HX} H^\dagger H X^\dagger X  + \lambda_{SX} S^\dagger S X^\dagger X.
\end{align}
The trilinear $\mu$ term allows a remnant discrete $Z_2$ symmetry after $S$ gets VEV and $\U1mt$ is spontaneously broken.
This local $Z_2$ symmetry~\cite{Krauss:1988zc} stabilises the DM candidate, which we assume the lighter component of $X$.
The kinetic mixing angle $\chi$ is strongly constrained to a level of $O(10^{-3})$ by the DM direct search
experiments~\cite{Mambrini:2011dw}. { The non-vanishing $\chi$ does not help solving
$b \to s\mu\mu$ anomaly because the SM gauge bosons allow only LFU couplings.}
 In this paper we neglect this term for simplicity
by setting $\chi \equiv 0$. 
We note that the fermion $Q$ which has the same SM quantum numbers with the left-handed quark doublets is vector-like under
both $\U1mt$ and the SM gauge groups.

We now consider the particle spectra and identify the DM candidate.
After { $H$ and $S$ get VEVs, $v_{H}$ and $v_S$}, the $\mu$ term makes the complex scalar $X$ split into two real scalar fields $X_{R,I}$ defined by
\begin{align}
  X \equiv {1 \over \sqrt{2}} (X_R + i X_I),
  \label{eq:X_RI}
\end{align}
with masses-squared
\begin{align}
 m_{R,I}^2 = m_X^2 +  {1 \over 2} \lambda_{HX} v_H^2 + {1 \over 2}\lambda_{SX} v_S^2  \pm \sqrt{2} \mu v_S.
\end{align}
Assuming $\mu >0$, the $X_I$ which is the lightest $Z_2$-odd neutral field is identified as the DM candidate.
The other $Z_2$ odd fields after $S$ gets VEV are $X_R$ and $Q$. The remaining particles including the SM fields are $Z_2$-even.
We take $m_{R,I}$, $\lambda_{HX}, \lambda_{SX}$ as free parameters, then we can write the parameters $m_X^2$ and $\mu$ in the Lagrangian as
\begin{align}
  m_X^2 &= \frac{m_R^2+m_I^2}{2} -{1 \over 2} \lambda_{HX} v_H^2 -{1 \over 2} \lambda_{SX} v_S^2, \nl
 \mu &= \frac{m_R^2-m_I^2}{2 \sqrt{2} v_S}.
\label{eq:mX_mu}
\end{align}
After $S$ gets VEV, the $\U1mt$ gauge boson $Z'$ obtains mass,
\begin{align}
m_{Z'} =  g_{Z'} |q_S| v_S   = 2  g_{Z'} |q_X| v_S, 
\end{align}
where $g_{Z'}$ is the gauge coupling constant of the $\U1mt$ group. The vector-like quark $Q$ does not mix with the ordinary quark
with the same SM quantum numbers because it is $Z_2$-odd while the SM counterparts are $Z_2$-even.
So it is already in the mass eigenstates with mass $M_Q$ at tree level, though the mass-splitting can be generated at loop-level.
This also distinguishes our model from the models in \cite{Belanger:2015nma, Altmannshofer:2014cfa}.

In the unitary gauge we decompose the SM Higgs $H$ and the dark scalar $S$ as
\begin{align}
H =\left( \begin{array}{c} 0 \\ {1 \over \sqrt{2}}(v_H +h) \end{array}\right), \quad
S = {1 \over \sqrt{2}} (v_S + s).
\end{align}
The stationary condition at the vacuum gives conditions
\begin{align}
  \mu_H^2 &= \lambda_H v_H^2 + {1 \over 2} \lambda_{HS} v_S^2, \nl
  \mu_S^2 &= \lambda_S v_S^2 + {1 \over 2} \lambda_{HS} v_H^2.
\end{align}
Using the above conditions it is straightforward to obtain the scalar mass-squared matrix
\begin{equation}
\left(
\begin{array}{cc}
2 \lambda_H v_H^2 & \lambda_{HS} v_H v_S \\
\lambda_{HS} v_H v_S & 2 \lambda_S v_S^2 \\
\end{array}
\right),
\end{equation}
in the basis $(h,s)$. It is diagonalised by introducing mixing angle $\alpha_H$ to get the scalar mass eigenstates $(H_1,H_2)$
\begin{align} 
\left(
\begin{array}{c}
h \\ s
\end{array}
\right)
=
\left(
\begin{array}{cc}
\cos\alpha_H & \sin\alpha_H\\
-\sin\alpha_H & \cos\alpha_H \\
\end{array}
\right)
\left(
\begin{array}{c}
H_1 \\ H_2
\end{array}
\right),
\end{align} 
where $H_1$ is identified with the SM-like Higgs boson with mass $m_{H_1}=125$ GeV.
We will take $m_{H_1}$, $m_{H_2}$, and $\alpha_H$ as input parameters. Then the parameters $\lambda_{H,S}$ and $\lambda_{HS}$ are
derived from them,
\begin{align}
  \lambda_H &=\frac{ m_{H_1}^2 c_H^2 + m_{H_2}^2 s_H^2}{2 v_H^2}, \nl
  \lambda_S &=\frac{ m_{H_1}^2 s_H^2 + m_{H_2}^2 s_H^2}{2 v_S^2},  \nl
  \lambda_{HS} &=\frac{ (m_{H_2}^2 - m_{H_1}^2)s_H c_H}{ v_H v_S}, 
\end{align}
where $c_H (s_H)$ is an abbreviation of $\cos\alpha_H (\sin\alpha_H)$. We require $\lambda_{HS} > -2 \sqrt{\lambda_H \lambda_S}$ to
stabilise the scalar potential at the electroweak (EW) scale.

In (\ref{eq:model}) we  assume that the down-type quarks are already in the mass basis and that the flavor mixing due to 
Cabibbo-Kobayashi-Maskawa (CKM) matrix $V$ appears in the up-quark sector, {\it i.e.} $d_{iL}=d'_{iL}, u_{iL} =\sum_j V_{ji}^* u'_{jL}$ where primes represent
the mass eigenstates. Then the Yukawa interactions of the type $q-Q-X$ can be written in the form
\begin{align}
  \Delta {\cal L}_{\rm Yukawa} &= -{1 \over \sqrt{2}} \sum_{i=1,2,3} \left(\lambda_{u_i}  \ol{u}_{iL}^{\prime} U
                                 +\lambda_{d_i} \ol{d}_{Li}^\prime D \right)(X_R +i X_I)+h.c.,
\end{align}
where $\lambda_{u_i} \equiv \sum_j V_{ij} \lambda_j$ and $\lambda_{d_i} \equiv \lambda_i$. We will also use the notation
$\lambda_{d,s,b}$ for $\lambda_{d_i}$  and $\lambda_{u,c,t}$ for $\lambda_{u_i} (i=1,2,3)$.
We will simply set $\lambda_1=0$ to remove the constraints related to the first generation quarks. Even in this case we see
\begin{align}
\lambda_u = V_{us} \lambda_2 + V_{ub} \lambda_3,
\end{align}
is induced. The induced $\lambda_u$ can generate NP contribution to $D^0-\ol{D}^0$ mixing.
However, due to Cabibbo-suppressed contribution to $D^0-\ol{D}^0$ at least by ${\cal O}(\lambda^2)$ { where
$\lambda (\approx  0.23)$ is the Cabibbo angle},
the constraint from $D^0-\ol{D}^0$ can be always
satisfied once the constraint from $B_s-\ol{B}_s$ is imposed~\cite{Arnan:2016cpy}.  And we do not consider this constraint further.

The DM interacts with the SM fields through the {\it Higgs-portal} Lagrangian
\begin{align}
  {\cal L}(X_I X_I H_{1,2})  =& -{1 \over 2} \Big[\lambda_{HX} v_H c_H- (\lambda_{SX} v_S-\sqrt{2} \mu) s_H \Big]  H_1 X_I^2 \nl
                                        &-{1 \over 2} \Big[\lambda_{HX} v_H s_H  + (\lambda_{SX} v_S-\sqrt{2} \mu) c_H \Big]  H_2 X_I^2.
\label{eq:HPcoupl}
\end{align}
In this paper we will set $\alpha_H=0$ and $\lambda_{HX}=0$ to suppress the stringent constraint from the dark matter direct detection
experiments via this Higgs portal interaction~\cite{Baek:2011aa,Baek:2012se}. 

\section{$R_K^{(*)}$ and constraints}
\label{sec:NP}

\begin{figure}[t]
\begin{center}
\includegraphics[width=.85\textwidth]{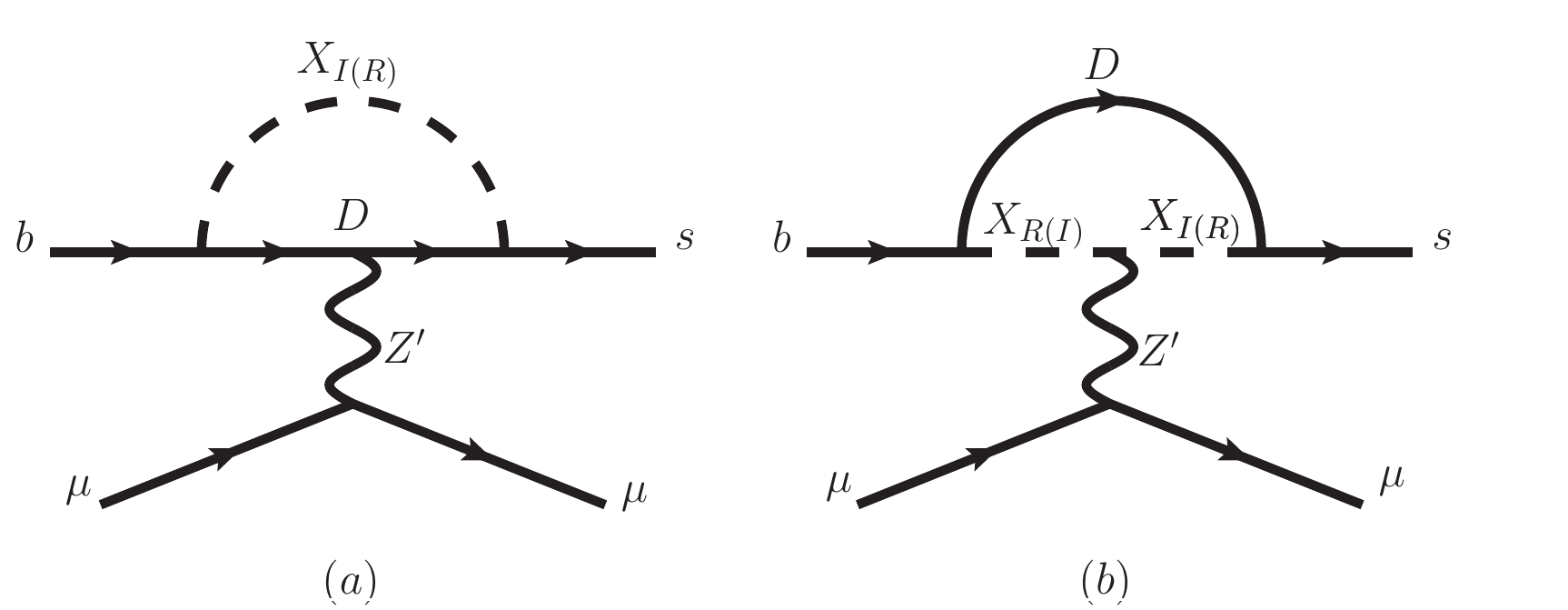}
\end{center}
\caption{$Z'$-exchanging penguin diagrams generating $b \to s \mu^+ \mu^-$ transition.}
\label{diag:C9}
\end{figure}
The $b \to s \mu^+ \mu^-$  transition operator $O_{9}^\mu$ which can explain the $R_{K^{(*)}}$ anomaly
is generated via the penguin diagrams shown in Fig.~\ref{diag:C9}. The arrows represent color or lepton number flow.
 First, we calculate the one-loop effective $s(p_s)-b(p_b)-Z'(q)$ ($q=p_s-p_b$) vertex.
 Assuming $m_I, m_R$, and $M_D$ are at the EW scale ($\equiv M_{\rm EW}$), we can neglect terms proportional to
external quark mass squareds, $m_{s(b)}^2/M_{\rm EW}^2 (\ll 1)$.
In this approximation, it is straightforward to get the effective vertex for diagrams (a) and (b) in Fig.~\ref{diag:C9}:
 \begin{align}
   i V_\mu^{(a)} &=  \frac{i q_Q g_{Z'} \lambda_s \lambda_b^* }{2 (4\pi)^2}\sum_{i=I,R}\Big[M_D^2 C_0^{(i)} + q^2 C_{12}^{(i)}
                   - (d-2)  C_{00}^{(i)} \Big] \ol{u}(p_s) \gamma_\mu P_L u(p_b), \nl
   i V_\mu^{(b)} &= \frac{-i q_X  g_{Z'}\lambda_s \lambda_b^* }{ (4\pi)^2} \Big[ C_{00}^{(IR)}+C_{00}^{(RI)} \Big] \ol{u}(p_s) \gamma_\mu P_L u(p_b),                   
 \end{align}
 where we take the dimension of space-time integration $d$ to be $d \equiv 4 -2\epsilon$ for positive infinitesimal $\epsilon$.
The $C$'s are abbreviations for one-loop three-point functions defined in~\cite{Hahn:1998yk},
 \begin{align}
   C_k^{(i)}&=C_k(m_s^2, q^2, m_b^2, m_i^2,M_D^2, M_D^2), \nl
   C_k^{(ij)}&=C_k(m_s^2, q^2, m_b^2, M_D^2, m_i^2, m_j^2), 
 \end{align}
 where $k=0, 12,00$ and $i,j=I,R (i \centernot = j)$. We will set $m_s=m_b\equiv 0$ in the calculation of the $C$-functions to be consistent with
 our approximation $m_{s(b)}^2/M_{\rm EW}^2 \ll 1$. The $C_{00}$-functions are divergent while $C_0$- and $C_{12}$-functions are finite.
 The divergence in the $C_{00}$-functions can be isolated as
 \begin{align}
   C_{00} ={1 \over \epsilon} -\gamma_E + \log 4 \pi +C_{00}|_{\rm finite},
 \end{align}
 where $C_{00}|_{\rm finite}$ is the remaining finite part. Using the relation between the $\U1mt$ charges, $q_Q = -q_X$,
 we can show that the sum of the two one-loop effective vertices
 is finite and given by
 \begin{align}
   i V_{\rm eff}^\mu (q^2) &\equiv i (V_\mu^{(a)}+V_\mu^{(b)}) \nl
       &= \frac{-i q_X g_{Z'} \lambda_s \lambda_b^* }{32 \pi^2} \, {\cal V}_{sb}(q^2,M_D^2, m_I^2, m_R^2) \,\ol{u}(p_s) \gamma_\mu P_L u(p_b),                   
 \end{align}
where
 \begin{align}
   {\cal V}_{sb}(q^2, M_D^2, m_I^2, m_R^2) &= 1+M_D^2 ( C_0^{(I)}    + C_0^{(R)}   ) + q^2 (C_{12}^{(I)} + C_{12}^{(R)})
   -2 (C_{00}^{(I)}|_{\rm  finite} +C_{00}^{(R)}|_{\rm finite}) \nl
   &    -2 (C_{00}^{(IR)}|_{\rm  finite} +C_{00}^{(RI)}|_{\rm finite}).
 \end{align}
 Now we can attach the external muon line in Fig.~\ref{diag:C9} to the $Z'$ to get $C_9^\mu$. The full amplitude for $b \to s \mu \mu$
 transition in Fig.~\ref{diag:C9} is given by
 \begin{align}
   i A &= -i g_{Z'} V_{\rm eff}^\mu(q^2) \frac{g_{\mu\nu}-q_\mu q_\nu/m_{Z'}^2}{q^2-m_{Z'}^2} \ol{u}(p_3) \gamma^\nu u(p_4),
 \end{align}
 where $p_3 (p_4)$ is outgoing (incoming) muon four-momentum. The term proportional to $q_\mu q_\nu$ vanishes because
 $\ol{u}(p_3) \centernot q u(p_4) =0$. Since $q \sim {\cal O}(m_b)$ at most, we can set $q^2 \equiv 0$ in the denominator of
 $Z'$-propagator.
 In this case the effective vertex can be written in a simple analytic form:
 \begin{align}
   {\cal V}_{sb}(0, M_D^2, m_I^2, m_R^2)&={1 \over 2}(k'(x_I)+k'(x_R))-k(x_I,x_R),
 \end{align}   
 where $x_{I(R)}=m_{I(R)}^2/M_D^2$ and the loop function $k$ is defined in Appendix~\ref{app:loop}.
We note ${\cal V}_{sb} \to 0$, when $x_R \to x_I$.
 This can be understood as follows: in the limit $m_R \to m_I$, the two real scalars $X_I$ and $X_R$ merge into the original complex scalar
 $X$ as can be seen from (\ref{eq:X_RI}). In this limit, the subset of the full Lagrangian which contributes the effective vertex 
$i V_{\rm eff}^\mu (q^2)$,
 \begin{align}
   \label{eq:L_sub}
\Delta {\cal L} &= \ol{D} (i \centernot D -M_D) D + D_\mu X^\dagger D_\mu X -m_X^2 X^\dagger X-(\lambda_b \ol{b}_L D X  + \lambda_s \ol{s}_L D X +{h.c.}),
 \end{align}
is invariant under local $\U1mt$ symmetry. Note that the $Z'$ mass term which breaks the $\U1mt$ is not included in $\Delta {\cal L}$.
Then the Ward-Takahashi identity dictates $V_{\rm eff}(q^2=0)=0$\footnote{This result holds also in case we keep the quark masses $m_s$ and $m_b$
because they respect the $\U1mt$ symmetry.}  due to the absence of the tree-level $Z'$-exchanging FCNC, leading to $V_{\rm eff}(q^2) \propto q^2$
to all orders of perturbation theory ~\cite{Peskin:1995ev}. Since $V_{\rm eff}(q^2=0)=0$, we obtain ${\cal V}_{sb}=0$ in the limits $q^2 \to 0$ and 
$ m_R \to m_I$.
\begin{figure}[t]
\begin{center}
\includegraphics[width=.35\textwidth]{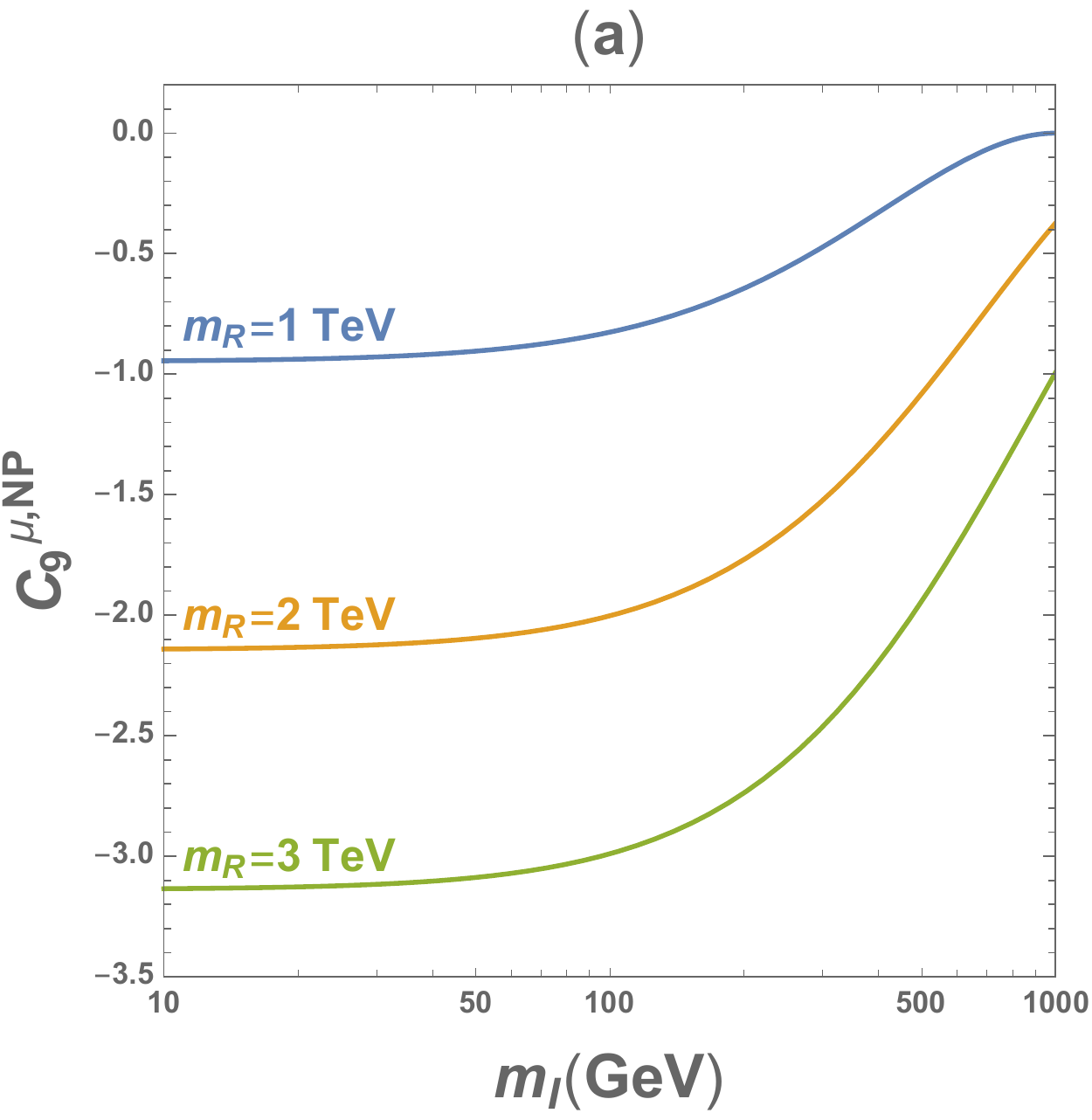}
\includegraphics[width=.35\textwidth]{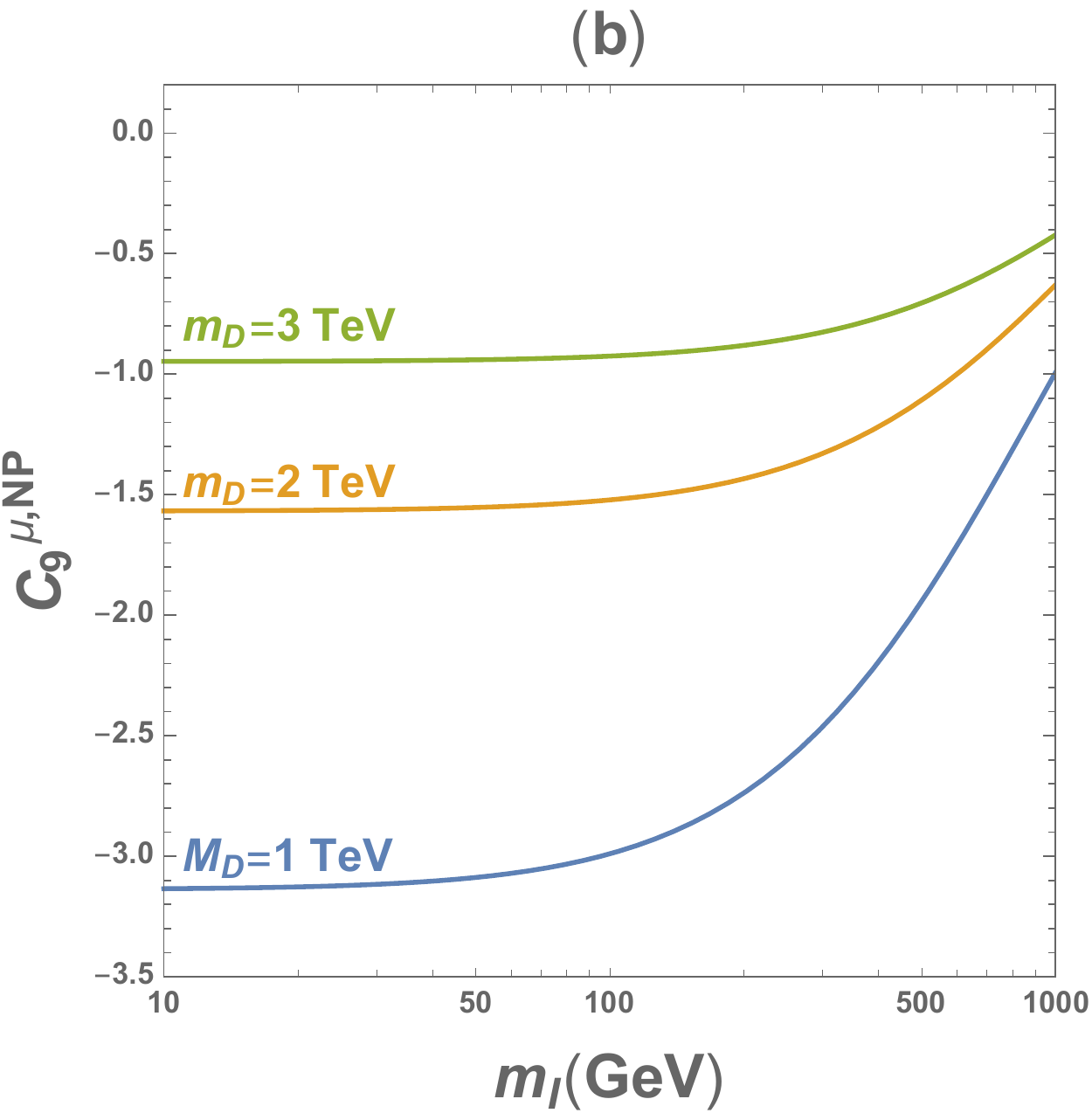}
\end{center}
\caption{Plots of $C_9^{\mu, {\rm NP}}$ as a function of $m_I$ for $m_R=1,2,3$ TeV (left panel) and for $M_D=1,2,3$ TeV (right panel). 
The fixed parameters are shown in the text.}
\label{fig:C9mI}
\end{figure}

 Now it is straightforward to get
 \begin{align}
   C_9^{\mu,{\rm NP}} = -\frac{\sqrt{2} q_X}{8 G_F m_{Z'}^2} \, \frac{\alpha_{Z'}}{\alpha_{\rm em}} \,\frac{\lambda_s \lambda_b^*}{ V_{ts}^*
   V_{tb} }   \Bigg[ {1 \over 2}(k'(x_I)+k'(x_R))-k(x_I,x_R)\Bigg],
\label{eq:C9NP}
 \end{align}
 where the prime on the $k$ functions denotes a derivative with respect to the argument and we fixed the $\U1mt$ charge of $\mu^-$ to be
 $+1$. A sizable mass splitting between $m_R$ and $m_I$ is favoured to generate $C_9^{\mu,{\rm NP}}$ which can explain the $\bsmm$ anomaly.
As a benchmark point in the parameter space, we choose $q_X=2$, $\alpha_{Z'}=0.1$, $m_{Z'}=700$ GeV, 
$\lambda_s \lambda_b^*=0.2$, $M_D=1$ TeV, $m_I=900$ GeV, and $m_R=3$ TeV, for which we get
\begin{align}
C_9^{\mu,{\rm NP}}  &=-1.14 \left(q_X \over 2\right)\left(\alpha_{Z'} \over 0.1\right)\left(\lambda_s \lambda_b^*  \over 0.2\right),
\label{eq:C9_benchmark}
\end{align}
which is close to the best fit value in (\ref{eq:C9_fit}) to solve the $b \to s\mu\mu$ anomaly. 
Fig.~\ref{fig:C9mI}(a) shows $C_9^{\mu,{\rm NP}}$ as a function of $m_I$ for three different values of $m_R=1,2,3$ TeV (from above)
with $m_{Z'}=700$ GeV, $M_D=1$ TeV, $q_X=2$, $\alpha_{Z'}=0.1$, and $\lambda_s\lambda_b^*=2$.
As can be seen from (\ref{eq:C9NP}),  $|C_9^{\mu,{\rm NP}}|$ becomes larger as the mass splitting $m_R-m_I$ increases.
Fig.~\ref{fig:C9mI}(b) shows a plot in the same plane but by varying $M_D=1,2,3$ TeV (from below)
with $M_R=3$ TeV and the other parameters the same as Fig.~\ref{fig:C9mI}(a).
We can see that the effect of the vector-like $D$-quark decouples as $M_D$ increases. In both cases, smaller $m_I$ is favored to
obtain larger $|C_9^{\mu,{\rm NP}}|$.

Photon- or $Z$-penguin diagrams similar to $Z'$-penguin diagrams in Fig.~\ref{diag:C9} but with
$Z'$ replaced by photon or $Z$-boson can contribute to $C_9^{\mu,{\rm NP}}$. Their couplings to leptons are flavour-universal and
they also contribute to $C_9^{e,{\rm NP}}$ and $C_9^{\tau,{\rm NP}}$ with the same value. So we use them as a constraint on the model.
The one-loop effective vertices they generate are proportional to $q^2$ by the same logic used to show $V_{\rm eff}(q^2) \propto q^2$
above. Here the conserved $U(1)$ symmetries are the $U(1)$-electromagnetism, $U(1)_{\rm em}$, for photon vertex, and the neutral current part of
$SU(2)_L$, $U(1)_Z$, for $Z$-boson vertex. Since these symmetries are conserved whether $\U1mt$ is conserved or not,
the argument applies even when $m_I \centernot = m_I$. 
If we attach the external muon lines, the $q^2$ in the photon-vertex cancels $q^2$ in the
photon propagator, whereas the one in the $Z$-vertex does not. As a consequence, the $Z$-penguin contribution is negligible because it
is proportional to $q^2/M_Z^2$ with $q^2 \sim m_b^2$.
We obtain the photon penguin contribution to be
\begin{align}
C_9^{\ell,{\rm NP}} =-\frac{\sqrt{2} e_d}{8 G_F} \frac{\lambda_s \lambda_b^*}{V_{ts}^* V_{tb}} {1 \over M_D^2} \left(Q_1(x_I)+Q_1(x_R)
  \right),
\label{eq:bsll}
\end{align}
where $x_{I(R)}=m_{I(R)}^2/M_D^2$ and the loop function $Q_1(x)$ is listed in  (\ref{eq:loop_fn_Q1}).
For the benchmark point $\lambda_s \lambda_b^*=0.2$,  $M_D=1$ TeV, $m_I=900$ GeV, and $m_R=3$ TeV,
we get
\begin{align}
C_9^{\ell,{\rm NP}} = -4.45 \times 10^{-3} \left(\lambda_s \lambda_b^* \over 0.2 \right),
\end{align}
which is about three orders of magnitude smaller than the $Z'$ contribution to $C_9^{\mu,{\rm NP}}$ in (\ref{eq:C9_benchmark}).
And we can neglect the photon- and Z-penguin contributions.

\begin{figure}[t]
\begin{center}
\includegraphics[width=0.7\textwidth]{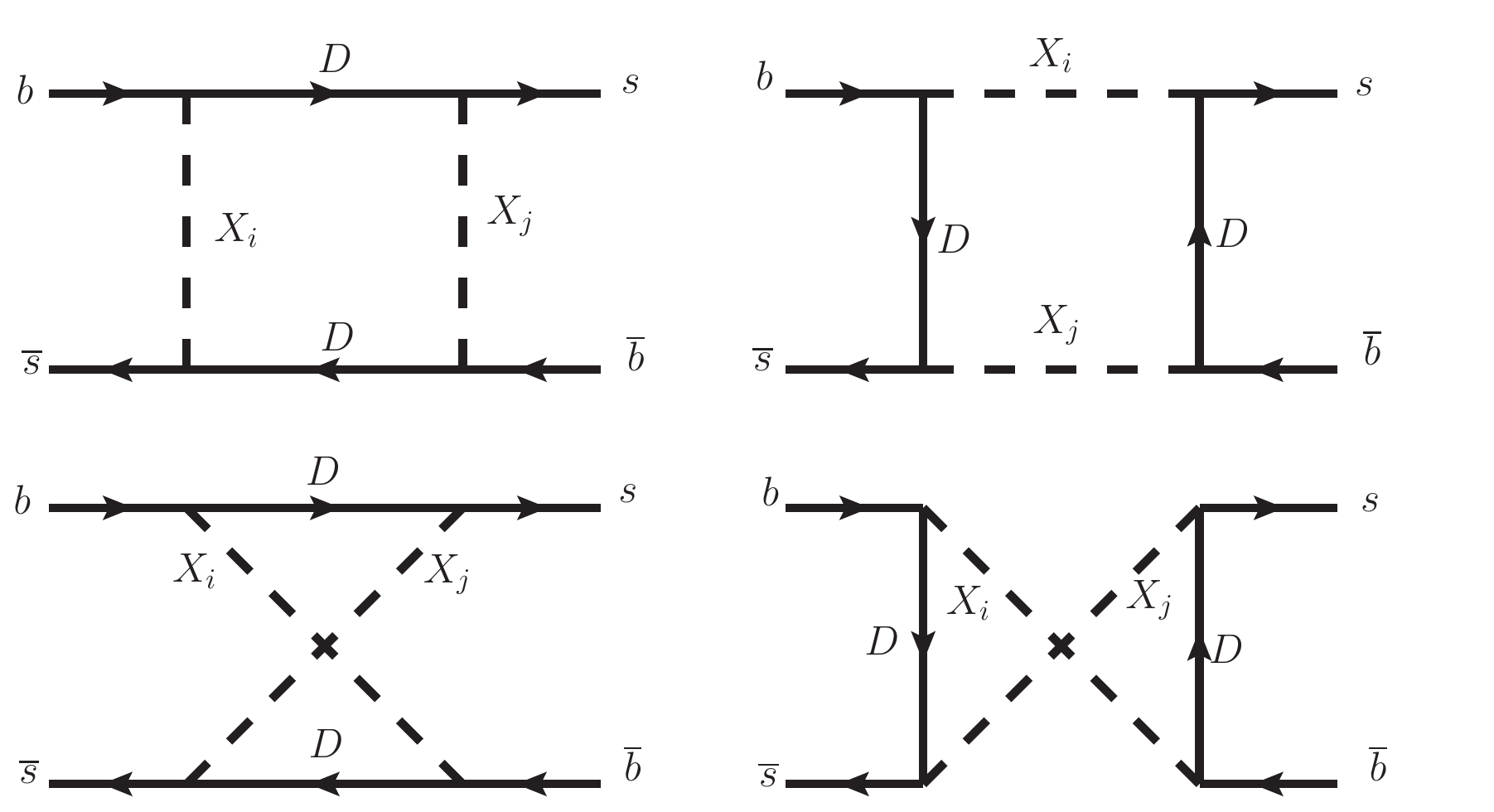}
\end{center}
\caption{Box diagrams for $B_s-\ol{B}_s$ mixing with $i,j=I,R$.}
\label{fig:BsBs}
\end{figure}

Now we consider other constraints on the model parameters. It turns out that the value $|\lambda_s \lambda_b^*|$
is the most strongly constrained by the measurements of the mass difference $\Delta m_s$ for $B_s-\ol{B}_s$ 
mixing.
Fig.~\ref{fig:BsBs} shows one-loop box diagrams for $B_s-\ol{B}_s$ mixing.
The arrows represent color flow. The lower two diagrams with crossed scalar lines exist because $X_I$ and $X_R$ are real scalars.
Our model where new particles couple only to the left-handed quarks contributes to the same effective operator with the one in the SM,
\begin{align}
{\cal H}_{\rm eff}^{\Delta B=2} &= C_1    (\ol{s} \gamma_\mu P_L b) (\ol{s}  \gamma^\mu P_L b). 
\end{align}
The Wilson coefficient $C_1$ can be decomposed into the SM and the NP contributions
\begin{equation}
C_1 = C_1^{\rm SM} + C_1^{\rm NP}.
\end{equation}
The SM contribution at the electroweak scale is obtained by box diagrams with $W$-boson and $t$-quark running inside the loop:
\begin{align}
C_1^{\rm SM} &= \frac{G_F^2 m_W^2}{4 \pi^2} (V_{ts}^* V_{tb})^2 S_0(x_t)  \approx 9.86 \times 10^{-11}\; {\rm GeV^{-2}},
\label{eq:BsBsSM}
\end{align}
where $x_t = m_t^2/m_W^2 \approx 4.64$ and the loop function $S_0(x_t)$ can be found in~\cite{Buchalla:1995vs}.
The NP diagrams shown in Fig.~\ref{fig:BsBs}  give
\begin{align}
C_1^{\rm NP} &= \frac{(\lambda_s \lambda_b^*)^2}{128 \pi^2 M_D^2} k(1,x_R,x_I),
\end{align}
where $x_i=m_i^2/M^2_D, \, (i=R,I)$. We note that this result is non-vanishing, different from a single real DM contribution which
vanishes~\cite{Arnan:2016cpy}. The non-zero term arises from the diagrams with $X_R$ and $X_I$ at the same time.
The measurement of the mass difference in the $B_s-\ol{B}_s$ system gives a constraint on the value of $C_1^{\rm NP}$: 
\begin{align}
-2.1 \times 10^{-11} \le C_1^{\rm NP} \le 0.6 \times 10^{-11} \, ({\rm GeV}^{-2}),
\label{eq:BsBs}
\end{align}
at 2$\sigma$ confidence level~\cite{Arnan:2016cpy}.
For the benchmark point $m_I=900$ GeV, $M_D=1$ TeV and $m_R=3$ TeV, we get
\begin{align}
C_1^{\rm NP} &= 0.473 \times 10^{-11} \left(\lambda_s \lambda_b^* \over 0.2 \right)^2\;{\rm GeV^{-2}},
\label{eq:C1NP_eg}
\end{align}
which is about an order of magnitude smaller than the SM prediction. This point satisfies the constraint (\ref{eq:BsBs}).

\begin{figure}[t]
\begin{center}
\includegraphics[width=.5\textwidth]{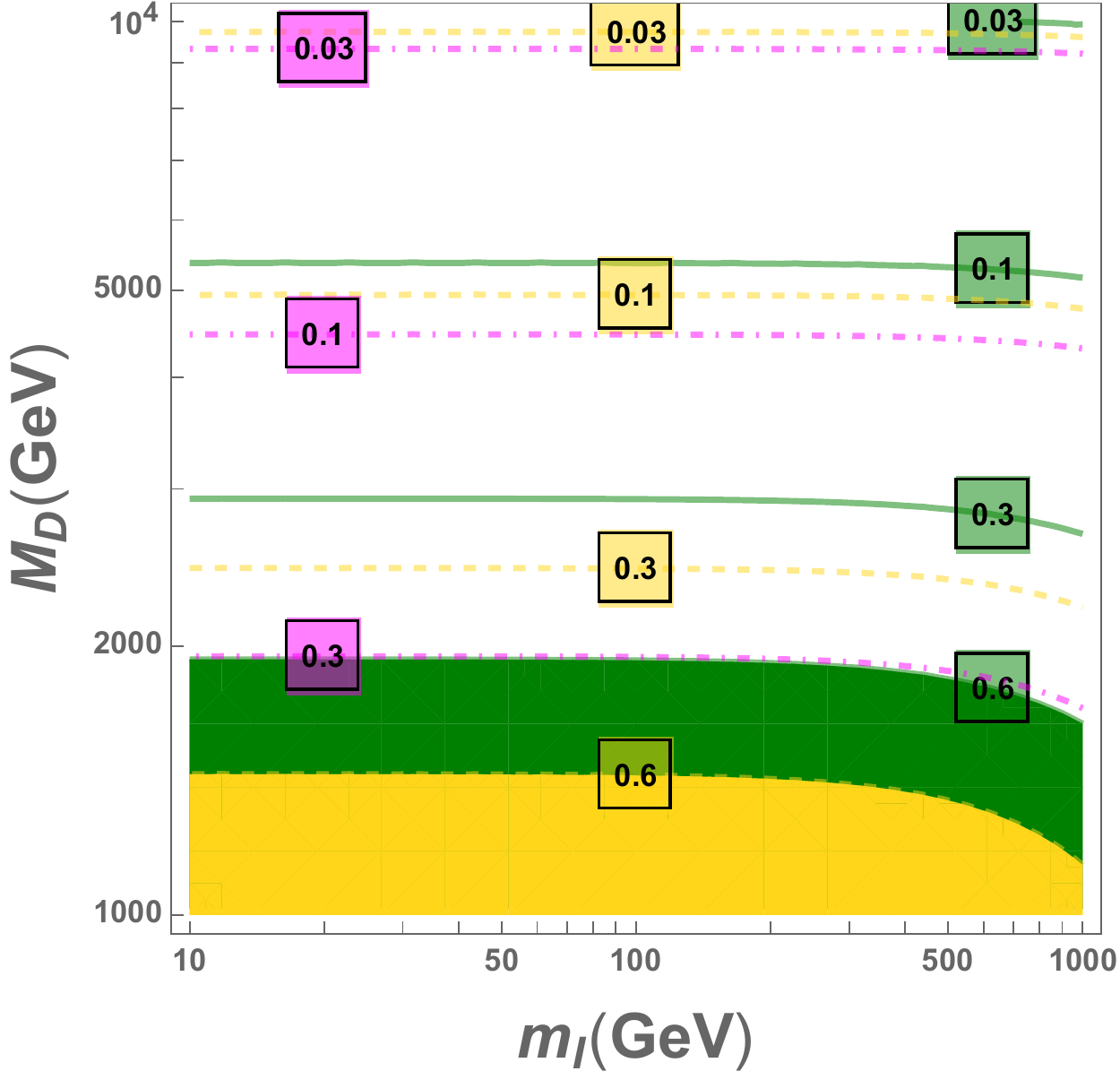}
\end{center}
\caption{A contour plot of $C_1^{\rm NP}$ with contour lines $0.03, 0.1, 0.3$, and $0.6$ in the unit of $10^{-11} \, {\rm GeV^{-2}}$  in the $(m_I, M_D)$-plane
for $\lambda_s \lambda_b^*=0.2$ and three different values of $m_R$. The  green solid (yellow dashed, magenta dot-dashed) lines correspond to $m_R=1 (2,3)$ TeV.
The green and yellow region is excluded for $m_R=1$ TeV and $m_R=2$ TeV, respectively. For $m_R=3$ TeV, the entire region is allowed.}
\label{fig:BsBs_plot}
\end{figure}

\begin{figure}[t]
\begin{center}
\includegraphics[width=.45\textwidth]{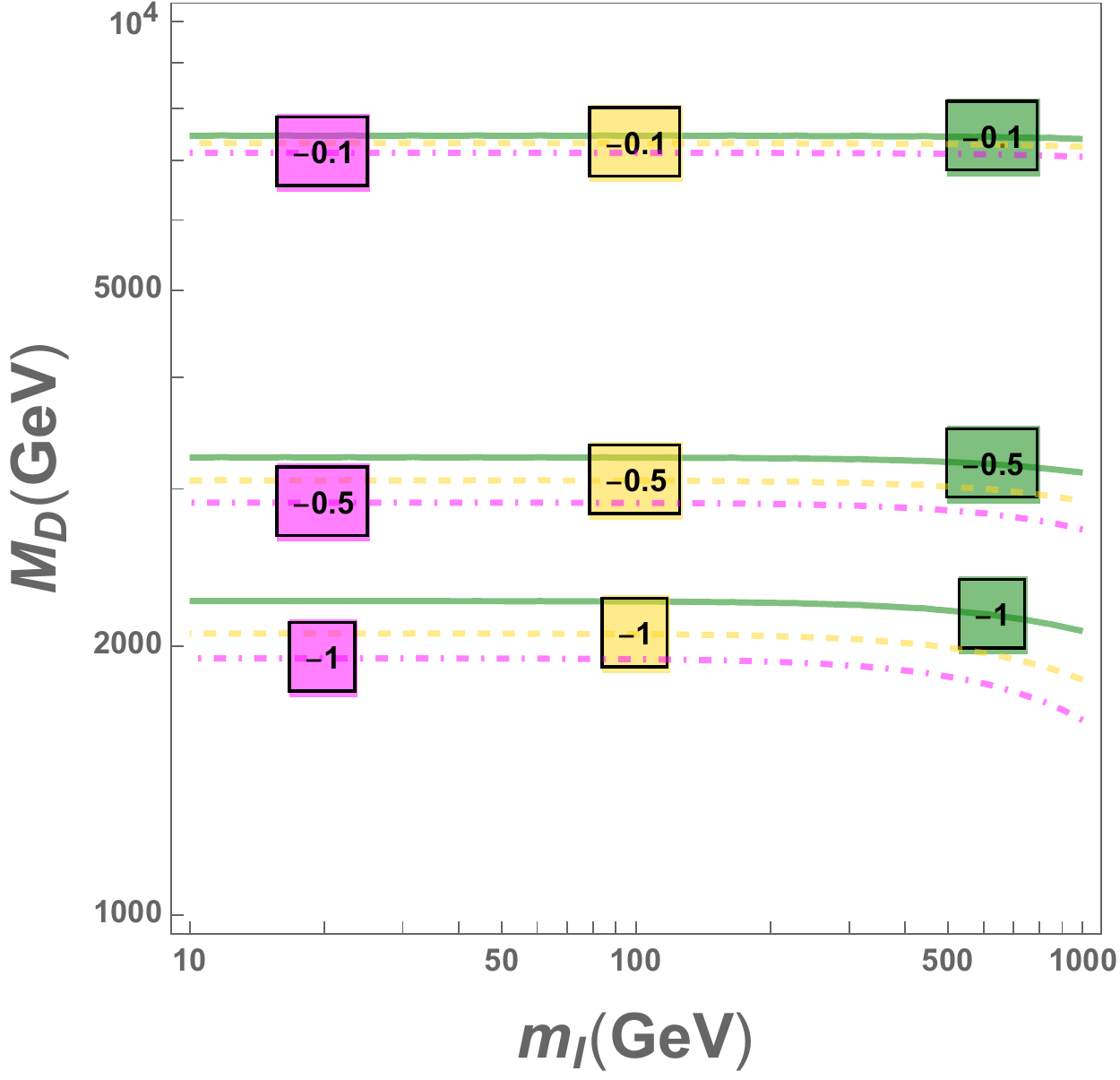}
\end{center}
\caption{A contour plot for the combination of Wilson coefficients $C_{7\gamma}+0.24\, C_{8g}$ with contour lines $-0.1,-0.5,-1.0$
  in the unit of $10^{-4}$  in the $(m_I, M_D)$-plane for $\lambda_s \lambda_b^*=0.2$ and three different values of $m_R$. 
The green solid (yellow dashed, magenta dot-dashed) lines correspond to  $m_R=1 (2,3)$ TeV.
}
\label{fig:bsr_plot}
\end{figure}

Fig.~\ref{fig:BsBs_plot} shows a contour plot of $C_1^{\rm NP}$ with
contour lines $0.03, 0.1, 0.3$, and $0.6$ in the unit of $10^{-11} \, {\rm GeV^{-2}}$  in the $(m_I, M_D)$-plane
for $\lambda_s \lambda_b^*=0.2$ and three different values of $m_R$. 
The green solid (yellow dashed, magenta dot-dashed) lines correspond to $m_R=1 (2,3)$ TeV. The green and yellow region is excluded by
(\ref{eq:BsBs}). The plot shows that $C_1^{\rm NP}$ is always positive in our model and the constraint (\ref{eq:C1NP_eg}) is easily satisfied when new
particles are at TeV scale.
\begin{figure}[t]
\begin{center}
\includegraphics[width=.4\textwidth]{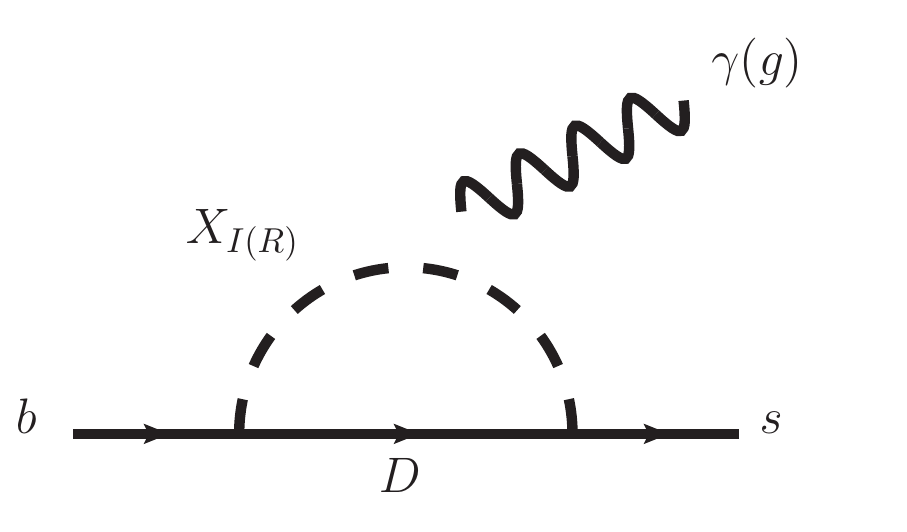}
\end{center}
\caption{Feynman diagrams for the new physics contributions to $b \to s \gamma  (g)$. The photon (gluon) line can be attached to any
  charged (colored) particles.}
\label{fig:bsr}
\end{figure}
Another possible constraint on the model parameters comes from the experimental measurements of the inclusive branching fraction of
radiative $B$-decay, $\ol{B} \to X_s \gamma$~\cite{Amhis:2016xyh},
\begin{align}
{\cal B}\left[\ol{B} \to X_s \gamma, \left(E_\gamma > 1.6 \,{\rm GeV}\right) \right]^{\rm exp} &= (3.32 \pm 0.16) \times 10^{-4}. 
\label{eq:bsr_exp}
\end{align}
For this process the SM prediction has been calculated up to NNLO QCD corrections~\cite{Misiak:2015xwa}, which predict,
\begin{align}
{\cal B}\left[\ol{B} \to X_s \gamma, \left(E_\gamma > 1.6 \,{\rm GeV}\right)\right]^{\rm SM} &= (3.36 \pm 0.23)  \times 10^{-4} .
\label{eq:bsr_SM}
\end{align}
The NP contribution to ${\cal B}(\ol{B} \to X_s \gamma)$ can be obtained by calculating the Wilson coefficients $C_{7\gamma, 8g}$ 
from the diagrams in Fig.~\ref{fig:bsr}: 
\begin{align}
C_{7\gamma}^{\rm NP}&= -\frac{\sqrt{2}}{16} e_D \frac{\lambda_s \lambda_b^*}{V_{ts}^* V_{tb}} \frac{1}{G_F M_D^2}
                      \left(J_1(x_I)+J_1(x_R)\right), \nl
C_{8g}^{\rm NP}&=  -\frac{\sqrt{2}}{16} \frac{\lambda_s \lambda_b^*}{V_{ts}^* V_{tb}} \frac{1}{G_F M_D^2}
                      \left(J_1(x_I)+J_1(x_R)\right), 
\end{align}
where $e_D=-1/3$ is the electric charge of the vector-like down-type quark $D$ and $x_i =m_i^2/M_D^2 \,(i=I,R)$.
The loop-fuction $J_1(x)$ is listed in the Appendix~\ref{app:loop}.
From the prediction including NP contribution to $C_{7\gamma(8g)}$~\cite{Misiak:2015xwa}, (\ref{eq:bsr_exp}) and (\ref{eq:bsr_SM}), we
obtain the constraint
\begin{align}
-6.3 \times 10^{-2} \le C_{7\gamma}^{\rm NP}+0.24\, C_{8g}^{\rm NP} \le 7.3 \times 10^{-2},
\label{eq:bsr_constraint}
\end{align}
at 2$\sigma$ level. For the benchmark point $m_I=900$ GeV, $M_D=1$ TeV, and $m_R=3$ TeV, we obtain
\begin{align}
C_{7\gamma}+0.24\, C_{8g} =-1.93 \times 10^{-4} \left(\lambda_q^2 \lambda_q^{3*} \over 0.2\right),
\end{align}
which is about two orders of magnitude less than the current bound (\ref{eq:bsr_constraint}). Fig.~\ref{fig:bsr_plot} shows a contour
plot for the combination $C_{7\gamma}+0.24\, C_{8g}$ with contour lines $-10^{-5},- 5 \times 10^{-5}$, and $-10^{-4}$ in 
the $(m_I, M_D)$-plane for $\lambda_s \lambda_b^*=0.2$ and three different values of $m_R$. 
The green solid (yellow dashed, magenta dot-dashed) lines correspond to  $m_R=1 (2,3)$ TeV.
We can see $C_{7\gamma}+0.24\, C_{8g}$ is less sensitive to
$m_R$ than $C_1^{\rm NP}$ of $B_s-\bar{B}_s$ is. The entire region considered is allowed by ${\cal B}^{\rm exp}(\ol{B} \to X_s \gamma)$.
The NP diagrams for semi-leptonic decay $B \to K^{(*)} \nu \bar{\nu}$ is obtained when the external muon lines are replaced with
neutrino lines in Fig.~\ref{diag:C9}. The effective Hamiltonian for the decay is
\begin{align}
{\cal H}_{\rm eff}^{\nu_i \nu_j} &= -\frac{4 G_F}{\sqrt{2}} V_{ts}^* V_{tb} C_L^{ij} O_L^{ij},
\end{align}
where
\begin{align}
O_L^{ij} = \frac{e^2}{16 \pi^2} (\ol{s} \gamma^\mu P_L b) (\ol{\nu}_i \gamma_\mu (1-\gamma_5) \nu_j).
\end{align}
We obtain the non-vanishing coefficients
 \begin{align}
   C_L^{22(33),{\rm NP}} = -(+)\frac{\sqrt{2} q_X}{16 G_F m_{Z'}^2} \, \frac{\alpha_{Z'}}{\alpha_{\rm em}} \,
\frac{\lambda_s \lambda_b^*}{ V_{ts}^*   V_{tb} }   \Bigg[ {1 \over 2}(k'(x_I)+k'(x_R))-k(x_I,x_R)\Bigg].
\label{eq:CnuNP}
 \end{align}
We note that the diagram with $Z'$ replaced by $Z$ vanishes in the $q^2 \to 0$ limit, showing that the $Z'$ contribution is dominant.  
The current experimental bounds on the ratios of branching fractions 
\begin{align}
R_{K^{(*)}}^{\nu\ol{\nu}} =\frac{{\cal B}(B \to K^{(*)} \nu \bar{\nu})^{\rm exp}}{{\cal B}(B \to K^{(*)} \nu \bar{\nu})^{\rm SM}}
\end{align}
are
\begin{align}
R_K^{\nu\ol{\nu}} < 4.3, \quad R_{K^*}^{\nu\ol{\nu}} < 4.4, \quad (\text{at 90\% C.L.}).
\end{align}
In our model these ratios are predicted to be
\begin{align} 
R_{K^{(*)}}^{\nu\ol{\nu}} &=\frac{\sum_{i,j=1}^3\left|C_L^{\rm SM} \delta^{ij} + C_L^{ij,{\rm NP}}\right|^2}{3 \left|C_L^{\rm SM} \right|^2}
=1+  \frac{2 \left|C_L^{22,{\rm NP}} \right|^2}{3 \left|C_L^{\rm SM} \right|^2},
\end{align} 
where we used $C_L^{33,{\rm NP}}=-C_L^{22,{\rm NP}} \centernot =0$ while all the other components are vanishing. We can see the interference
terms  cancel each other out.
Considering $C_L^{22,{\rm NP}}=C_9^{\mu,{\rm NP}}/2 \approx -0.6$  to explain the $b \to s\mu\mu$ anomaly
and $C_L^{\rm SM} \approx -6.35$, we predict
\begin{align}
R_{K^{(*)}}^{\nu\ol{\nu}}-1 \approx 7 \times 10^{-3}, 
\end{align}
showing the deviation from the SM is very small partly due to the cancellation of the interference terms.

\begin{figure}[t]
\begin{center}
\includegraphics[width=.4\textwidth]{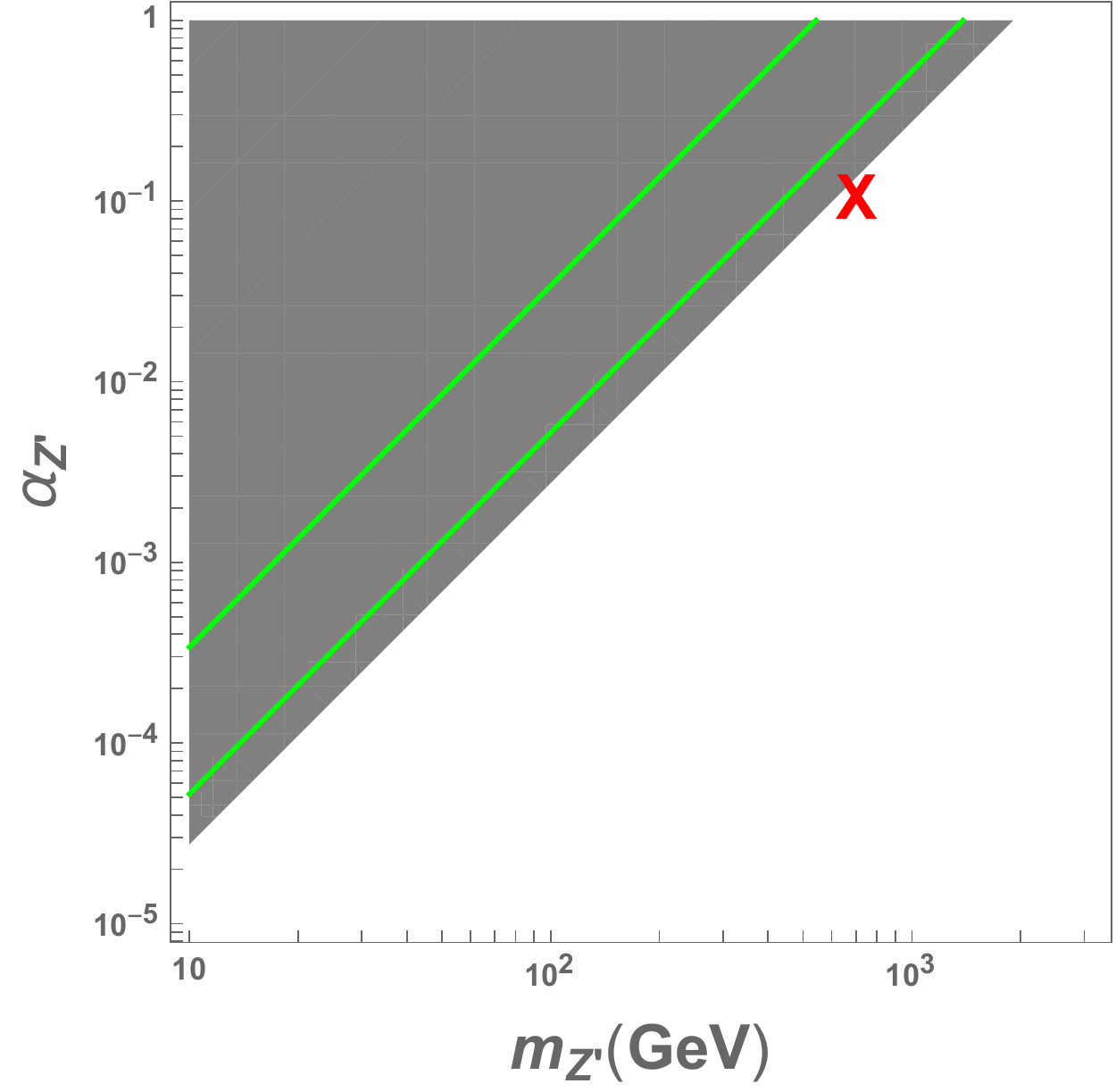}
\end{center}
\caption{The constraint from neutrino trident production and muon $g-2$ in $(m_{Z'},\alpha_{Z'})$ plane. The grey region is disfavoured
by the neutrino trident production experiments at 2$\sigma$. { The region between the two green lines is favoured by the current 
discrepancy $\Delta a_\mu$ at 2$\sigma$, but it is excluded by the neutrino trident production experiments.} 
The red ``X'' mark represents the benchmark point $m_{Z'}=700$ GeV 
and $\alpha_{Z'}=0.1$.}
\label{fig:trident}
\end{figure}

\begin{figure}[t]
\begin{center}
\includegraphics[width=.45\textwidth]{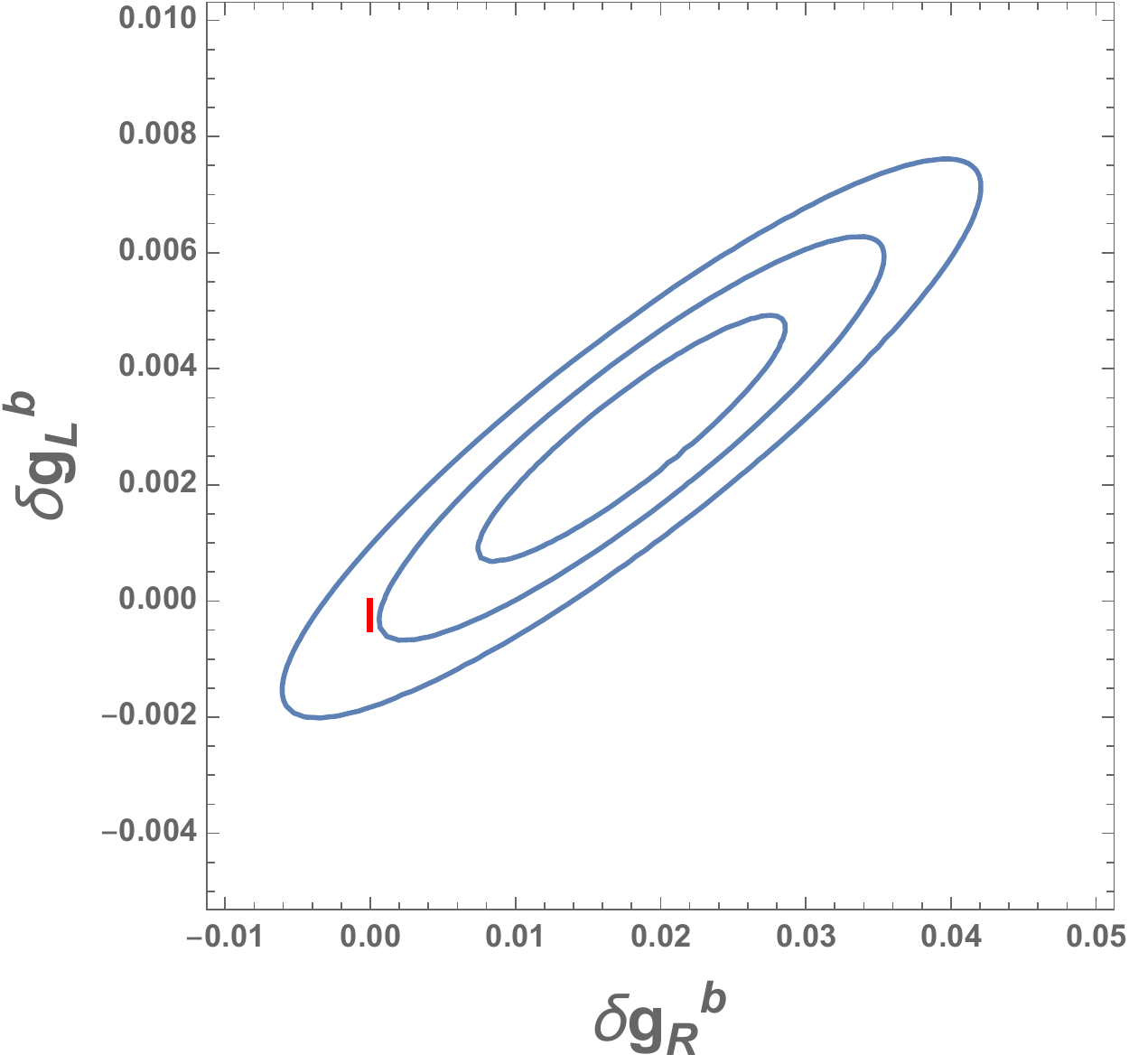}
\end{center}
\caption{The error ellipses at 1, 2, and 3$\sigma$ confidence level (from inside) in $(\delta g_R^b, \delta g_L^b)$ plane.
The thick red line segment represents the model prediction.}
\label{fig:gLR}
\end{figure}

The gauged $\U1mt$ model is well known to generate  the sizable anomalous magnetic moment of muon, $a_\mu=(g-2)_\mu/2$ via the
$Z'$-exchanging one-loop diagram~\cite{Baek:2001kca}.
The $Z'$ contribution can explain the long-standing discrepancy between the experimental measurements~\cite{Bennett:2006fi} and the SM
predictions~\cite{Kurz:2014wya}:
\begin{align}
\Delta a_\mu &=a_\mu^{\rm exp}-a_\mu^{\rm SM} = (236 \pm 87) \times 10^{-11}.
\label{eq:amu_diff}
\end{align}
The effective Hamiltonian for $a_\mu$ is
\begin{align}
{\cal H}^{a_\mu}_{\rm eff} &=-a_\mu  \frac{e}{4 m_\mu} (\ol{\mu} \sigma^{\mu\nu} \mu) F_{\mu\nu}.
\label{eq:amu_H}
\end{align}
The NP contribution at one-loop level is calculated to be
\begin{align}
a_\mu^{\rm NP} &= \frac{\alpha_{Z'}}{2\pi}\int_0^1 dx \frac{2 m_\mu^2 x^2 (1-x)}{x^2 m_\mu^2+(1-x) m_{Z'}^2}, 
\label{eq:amu_NP}
\end{align}
which in the limit, $m_\mu^2 \ll m_{Z'}^2$, approximates
\begin{align}
a_\mu^{\rm NP} \approx    \frac{\alpha_{Z'}}{3\pi} \frac{ m_\mu^2}{ m_{Z'}^2}.
\end{align}
For the benchmark point $m_{Z'}=700$ GeV, $\alpha_{Z'}=0.1$, we get
\begin{align}
a_\mu^{\rm NP} = 24.2 \times 10^{-11},
\end{align}
which is consistent with (\ref{eq:amu_diff}) within 3$\sigma$.
In the minimal $\U1mt$ model the region in the $(m_{Z'},\alpha_{Z'})$ plane which can explain $\Delta a_\mu$ at 2$\sigma$ level is
excluded by the bound from the measurement of neutrino trident production, $\nu_\mu N \to \nu_\mu N \mu^+ \mu^-$, when 
$m_{Z'} \gtrsim 400$ MeV~\cite{Altmannshofer:2014pba}. 
Fig.~\ref{fig:trident} shows the constraint from neutrino trident production and muon $g-2$ in $(m_{Z'},\alpha_{Z'})$ plane. 
The grey region is disfavoured by the neutrino trident production experiments at 2$\sigma$. 
{ The region between the two green lines is favoured by the current 
discrepancy $\Delta a_\mu$ at 2$\sigma$, but it is excluded by the neutrino trident production experiments.} 
 The red ``X'' mark represents the benchmark point $m_{Z'}=700$ GeV 
and $\alpha_{Z'}=0.1$.
The new particles in the model also generates one-loop effective $Z f \bar{f}$-vertex $(f=s,b)$.  Since $Z b \bar{b}$ vertex
has been more precisely determined by the LEP experiment, we consider the constraint only from $Z b \overline{b}$.
The $Z b \bar{b}$ vertex is written in the form,
\begin{align}
\Delta {\cal L}=-\frac{g}{\cos\theta_W} Z_\mu \ol{b} \gamma^\mu \left(g_L^b P_L + g_R^b P_R\right) b,
\end{align}
where tree-leve values for the couplings are  $g_L^b({\rm tree}) = -1/2 -e_d {s_w}^2$, $g_R^b({\rm tree}) = -e_d {s_w}^2$, 
$e_d=-1/3$, and ${s_w}^2 \approx 0.23$.
The deviation of $g_{R(L)}^b$ from the SM prediction obtained from a global fit is~\cite{Ciuchini:2013pca}\footnote{We choose
more conservative result in~\cite{Ciuchini:2013pca}.}
\begin{align}
\delta g_R^b = 0.018 \pm 0.007, \quad \delta g_L^b = 0.0028 \pm 0.0014, 
\label{eq:gLR_fit}
\end{align}
with a correlation coefficient of $+0.9$.
In our model the NP contributions to $g_L^b$ is obtained to be\footnote{The new particles couple to only $b_L$ and do not
generate $g_R^b$.},
\begin{align} 
g_L^{b,{\rm NP}} (q^2)=\frac{|\lambda_b^2|^2 q^2}{32 \pi^2 M^2_D} \left(-{1 \over 2}-sw^2 e_d\right)\left(Q_1(x_I)+Q_1(x_R)\right),
\end{align} 
where $x_{I(R)}=m_{I(R)}^2/M_D^2$ and we can take $q^2=m_Z^2$. 
The loop function $Q_1(x)$ is listed in (\ref{eq:loop_fn_Q1}). We notice that the loop function is the same with the one
for the photon penguin diagram of $b \to s \ell \ell$ in (\ref{eq:bsll}). In both cases the gauge bosons couple only to $D$
and the amplitudes are proportional to $q^2$ by Ward-Takahashi identity as we mentioned above (\ref{eq:bsll}). So they should be
proportional to each other.
The above result can be compared with (\ref{eq:gLR_fit}). In Fig.~\ref{fig:gLR} we show error ellipses at 1, 2, and 3$\sigma$
confidence level in $(\delta g_R^b, \delta g_L^b)$ plane. The vertical red line segment is obtained by randomly scanning $g_L^{b,{\rm NP}}$
in the ranges
\begin{align}
10^{-3} &< \lambda_b < 1, \nl
10 \, {\rm GeV}  &< m_I < 3 \, {\rm TeV}, \nl
m_I   &< m_R, M_D < 10 \, {\rm TeV}.
\end{align}
The model predicts $\delta g_L^{b,{\rm NP}}$ in the range $(-4.9 \times 10^{-4},0)$ and satisfies (\ref{eq:gLR_fit}) at 3$\sigma$ level.

The new particle searches at the LHC can also constrain the model. For example, new coloured-scalars $D$ or $U$ can be pair-produced
via $p p \to D \bar{D} (U \bar{U})$ at the LHC if their masses are within the LHC reach. These production processes are similar to
those considered in~\cite{Baek:2016lnv,Baek:2017ykw} where they were analysed in detail.
Roughly $M_{D(U)} \lesssim 1$ TeV are excluded. And we impose $M_D \ge 1$ TeV.

\section{The dark matter}
\label{sec:DM}
In this section we identify the main channel and the favoured parameter region to give the observed DM relic density, 
$\Omega_{\rm DM} h^2 =0.1199 \pm 0.0022$~\cite{Ade:2015xua}. 
We assume the weakly interacting massive particle (WIMP), $X_I$,  whose mass is at the electroweak scale,
 is the candidate for a cold dark matter (CDM) and constitute the
whole dark matter component in the universe. In addition we assume the DM relic came from the thermal freeze-out mechanism.
 In this mechanism, when they are at the initial equilibrium state for the high temperature, $T \sim m_I$,  the  DM particles whose number density
  is similar to that of the photon are overabundant.
The DM number density becomes reduced by (co)annihilations until their rates are smaller than the Hubble expansion rate,
when it freezes out typically near $T \sim m_I/25$~\cite{Kolb:1990vq}. 
Then the relic density is roughly related with the annihilation cross section at
freeze-out temperature as
\begin{align}
\Omega_{\rm DM} h^2 \approx \frac{3 \times 10^{-27} \, \mathrm{cm^3 /s}}{\langle \sigma_\mathrm{ann} v\rangle},
\end{align}
where $v$ is the relative velocity between the DM particles.

Before studying DM phenomenology, we can get insight by comparing our model with the minimal 
``scalar singlet dark matter'' model with $Z_2$ symmetry~\cite{Cline:2013gha}.
The scalar potential in the minimal model has terms
\begin{align}
V = {1 \over 2} \mu_D^2 D^2 + {1 \over 2} \lambda_{hD} D^2 |H|^2.
\end{align}
The DM mass $m_D$ is obtained by $m_D^2=\mu_D^2+\lambda_{hD} v_H^2/2$. The DM annihilation occurs through 
$D D \to h \to  \mathrm{SM} \,\mathrm{SM}$ or $D D \to h h$. Both processes are controlled by the Higgs portal
coupling $\lambda_{hD}$, which is strongly restricted by the direct detection experiments ~\cite{Akerib:2016vxi, Cui:2017nnn, Aprile:2018dbl}.
As a consequence the model is strongly constrained,
ruling out $m_D \lesssim 1$ TeV region as a single-component DM~\cite{Athron:2017kgt}.

\begin{figure}[t]
\begin{center}
\includegraphics[width=.9\textwidth]{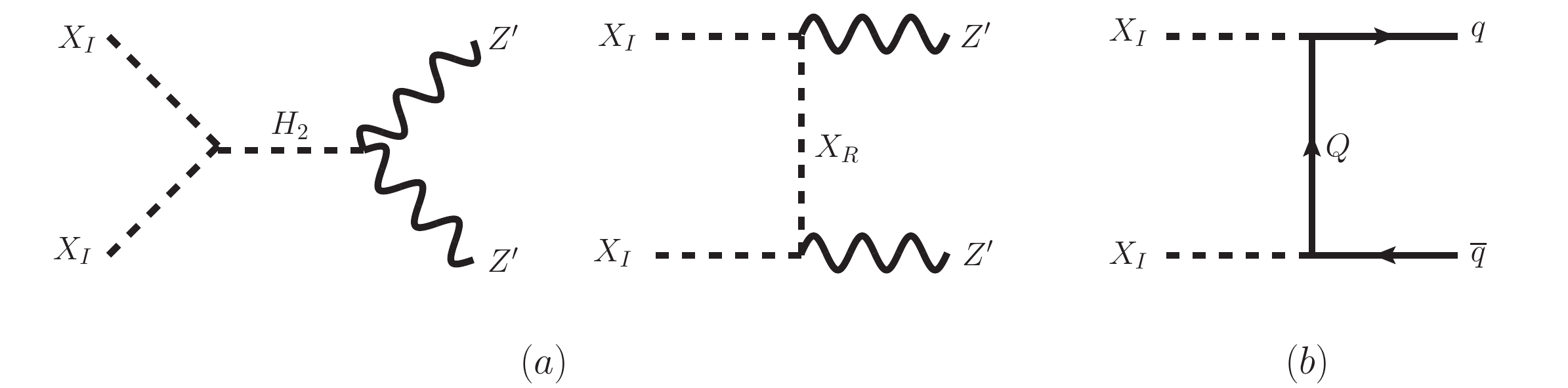}
\end{center}
\caption{Representative diagrams for the two annihilation channels: (a) $X_I X_I \to Z' Z'$ and (b) $X_I X_I \to q \bar{q}$ 
($q=u,d,s,c,b,t$).
}
\label{fig:relic}
\end{figure}

In our model, however, there are many model parameters
involved in the DM-Higgs couplings as can be seen in (\ref{eq:HPcoupl}), which allows the direct detection constraint on
the Higgs
portal interaction  to be
lifted by setting $\alpha_H=\lambda_{HX}=0$ to remove $H_1 X_I^2$ term. Even in that case the heavy Higgs $H_2$ can still
mediate the DM interaction without much affecting the DM scattering off nuclei. There are also dark gauge interaction and
dark Yukawa couplings available for DM annihilations.
In this paper we consider two processes for DM annihilation which can occur in different regions of parameter space:
$X_I X_I \to Z' Z'$ and $X_I X_I \to q \ol{q}$. Barring the Higgs portal $X_I$ interaction with $H_1$, they are dominant processes.
In Fig.~\ref{fig:relic} we show representative diagrams for the two annihilation channels. We implemented our
model to the micrOMEGAs package~\cite{Belanger:2013oya} to evaluate the DM relic density and direct detection cross section.

We first consider the scenario in which the diagrams of type (a) in Fig.~\ref{fig:relic} play a major role.
The process $X_I X_I \to Z' Z'$ dominates the DM annihilations
as long as it is kinematically open, $\alpha_{Z'}$ is not too small ($\alpha_{Z'} \gtrsim 10^{-6}$), and $m_{H_2}$ is not much larger
than TeV scale ($m_{H_2} \lesssim 5$ TeV). In this case we obtain typically 
$\langle \sigma v (X_I X_I \to Z' Z') \rangle \gg \langle \sigma v (X_I X_I \to q \bar{q}') \rangle$ for the thermal-averaged
annihilation cross sections.
Given that we set $\alpha_H=\lambda_{HX}=0$, the process $X_I X_I \to Z' Z'$ is controlled by the dark Higgs interaction 
and the dark gauge interaction.

\begin{figure}[t]
\begin{center}
\includegraphics[width=.38\textwidth]{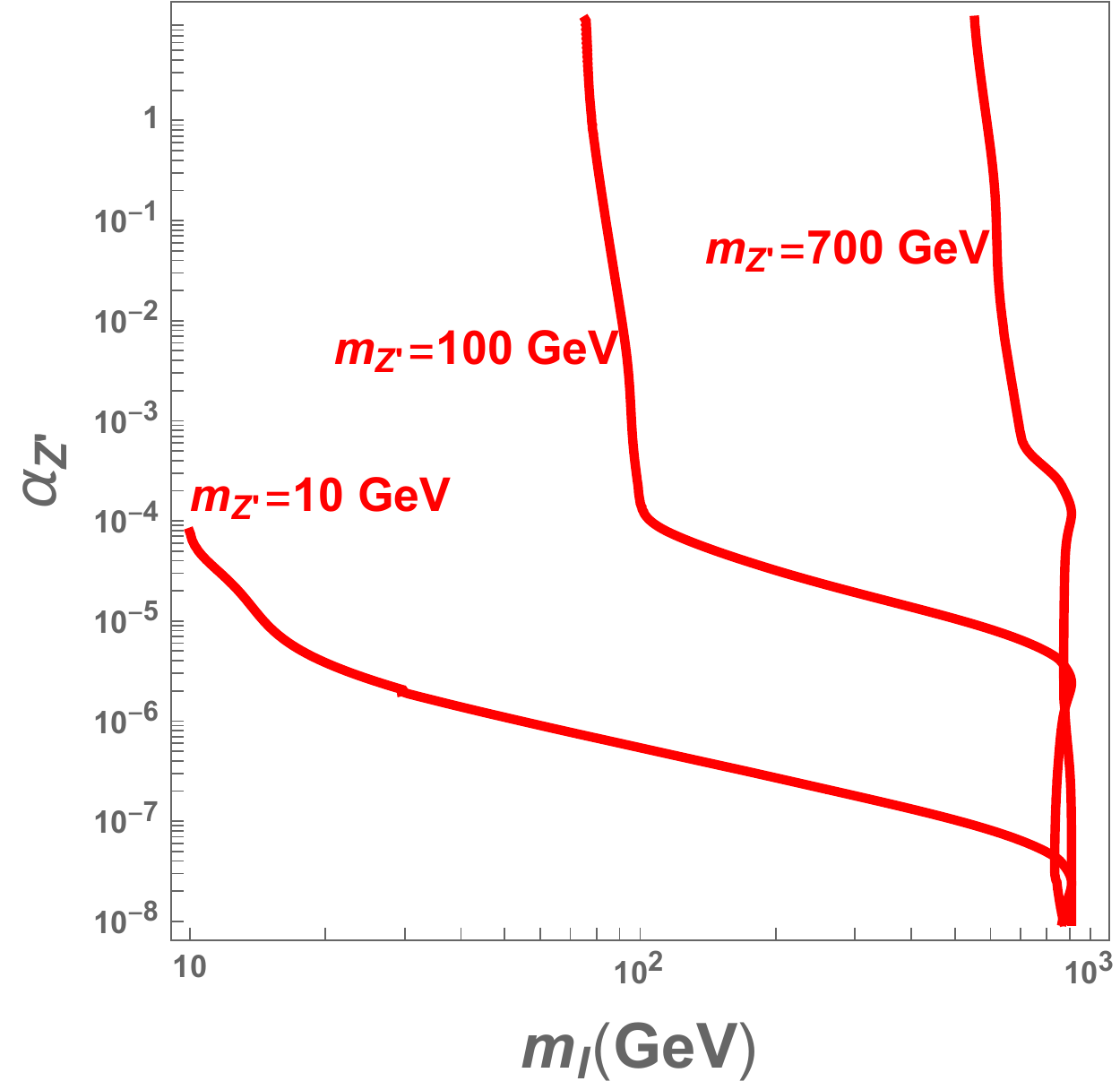}
\includegraphics[width=.38\textwidth]{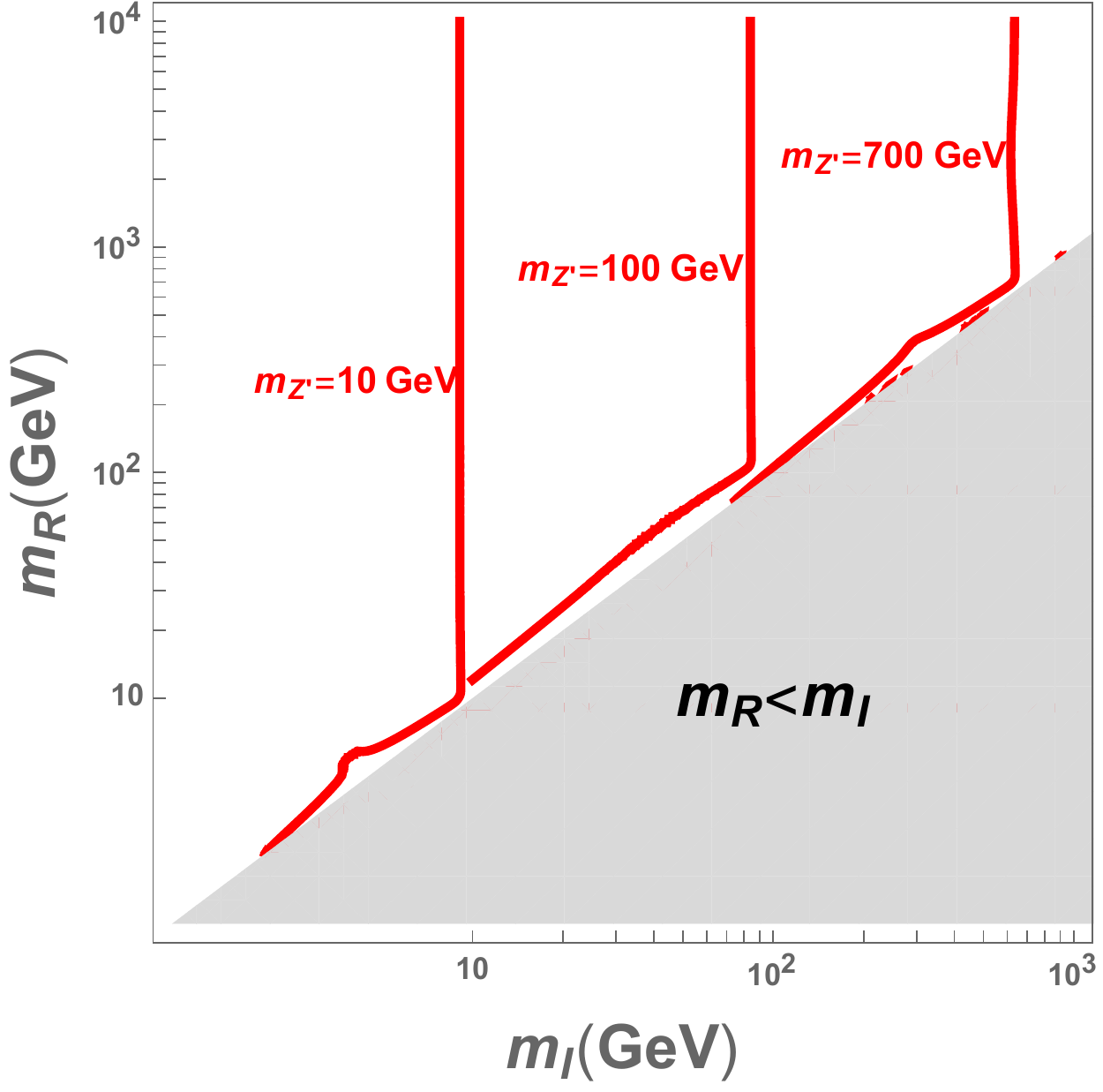}
\end{center}
\caption{Contours of $\Omega_{\rm DM} h^2=0.1199$ in $(m_I,\alpha_{Z'})$ (left panel) and $(m_I,m_R)$ (right panel) plane  for 
$m_{Z'}=10, 100, 700$ GeV (from left). { We set $m_R=3$ TeV for the left panel and $\alpha_{Z'}=0.1$ for the right panel.}
The other fixed parameters can be found in the text.
The grey region in the right panel gives $m_R < m_I$ and is 
not considered.}
\label{fig:MI-alphaX}
\end{figure}

The former interaction is  given by $\lambda_S v_S -\sqrt{2} \mu$, and the latter  by $g_{Z'} v_S$.
Both are sensitive to $C_9^{\mu,{\rm NP}}$ in (\ref{eq:C9NP}). 
Fig.~\ref{fig:MI-alphaX} shows contour lines of $\Omega_{\rm DM} h^2=0.1199$ in $(m_I,\alpha_{Z'})$ (left panel) 
and $(m_I,m_R)$ (right panel) plane for $m_{Z'}=10, 100, 700$ GeV. 
For the other parameters we take the benchmark values: $q_X=2$, $\lambda_s \lambda_b^*=0.2$,   $M_D=1$ TeV, 
and $m_{H_2}=2$ TeV. We set $m_R=3$ TeV for the left panel and $\alpha_{Z'}=0.1$ for the right panel.
 We also fixed $\alpha_H =0$, $\lambda_X=1$, $\lambda_{HX}=0$, and $\lambda_{SX}=0.2$.
At this stage we do not impose constraints other than $m_R > m_I$.
Numerically we have checked that $\langle \sigma v (X_I X_I \to Z' Z') \rangle$ is much larger than 
$\langle \sigma v (X_I X_I \to q \bar{q}') \rangle$ for the points in Fig.~\ref{fig:MI-alphaX}.

The process $X_I X_I  \to Z' Z'$ can occur in the early universe even when $m_I < m_{Z'}$ if the mass difference is not too large.
This can be seen in the steep lines corresponding to $m_I < m_{Z'}$ in the left panel of Fig.~\ref{fig:MI-alphaX}. 
This is possible when the DMs move fast and their center of mass energy exceeds twice the $m_{Z'}$: $\sqrt{s} > 2 m_{Z'}$.
To produce on-shell $Z'$-pair, the relative velocity of DM pair in the CM-frame should satisfy
\begin{align}
v \ge 2 \sqrt{1 - {m_I^2 \over m_{Z'}^2}}.
\end{align}
For example, for $m_I=90$ GeV and $m_{Z'}=100$ GeV, we obtain $v \ge 0.87$. The DM should be quite relativistic  and 
the thermally averaged annhilation cross section
is Boltzmann-suppressed.
When $X_I X_I \to Z' Z'$ is kinematically open for non-relativistic $X_I$, the process is sensitive to the dark gauge coupling
$\alpha_{Z'}$. For our benchmark point it turns out that the $H_2$-exchanging $s-$channel diagram is more important than the
$X_R$-exchanging $t-$channel diagram due to the $\mu-$term in (\ref{eq:HPcoupl}). This shows that the process is also
sensitive to the mass-squared difference, $m_R^2-m_I^2$, by (\ref{eq:mX_mu}).
When $m_I$ is not close to the resonance region, the $s$-wave annihilation cross section for the $X_I X_I \to H_2 \to Z' Z'$ channel is
in the form
\begin{align}
\sigma v 
=\frac{(8 \pi q_X^2 \alpha_X (m_R^2-m_I^2)-\lambda_{SX} m_{Z'}^2)^2(4m_I^4-4 m_I^2 m_{Z'}^2 + 3 m_{Z'}^4)\sqrt{m_I^2-m_{Z'}^2}}{16 \pi m_I^3 m_{Z'}^4 (m_{H_2}^2-4 m_I^2)^2} + {\cal O}(v^2).
\end{align}
When $\lambda_{SX}$ is not too large, the larger mass squared difference $m_R^2-m_I^2$ and the smaller $m_{Z'}^2$, {\it i.e.} the larger
$\mu$, the larger $\sigma v$ is obtained.

As $m_I$ approaches 1 TeV, it is close to the resonance region { $m_{H_2} \approx 2 m_I$} and the cross section increases rapidly,
{virtually} independent of $\alpha_{Z'}$. This explains almost vertical parts of the curves near $m_I = 1$ TeV. 

In the right panel of Fig.~\ref{fig:MI-alphaX}, we can see that the regions $m_R \approx m_I$ also give the correct relic density.
This occurs due to the coannihilation processes $X_R X_I \to Z' H_i (q \bar{q}, \ell \bar{\ell})$~\cite{Baek:2017sew}. 
As we saw in (\ref{eq:C9NP}), the NP contribution to $C_9^{\mu,{\rm NP}}$ is suppressed. And the coannihilation mechanism
for the DM relic density
is not favoured as a solution to $\bsmm$ anomaly. This shows a strong interplay between the flavour physics and DM 
phenomena~\cite{Baek:2001nz,Baek:2004et,Baek:2005wi,Baek:2005di,Baek:2015fma,Baek:2016kud,Baek:2017sew,Baek:2017ykw},
which will be discussed in the next Section in more detail.
Once $X_I X_I \to Z' Z'$ is
kinematically open near $m_I= m_{Z'}$, it dominates the annihilation processes, which is not so sensitive to $m_R$.

\begin{figure}[h]
\begin{center}
\includegraphics[width=.43\textwidth]{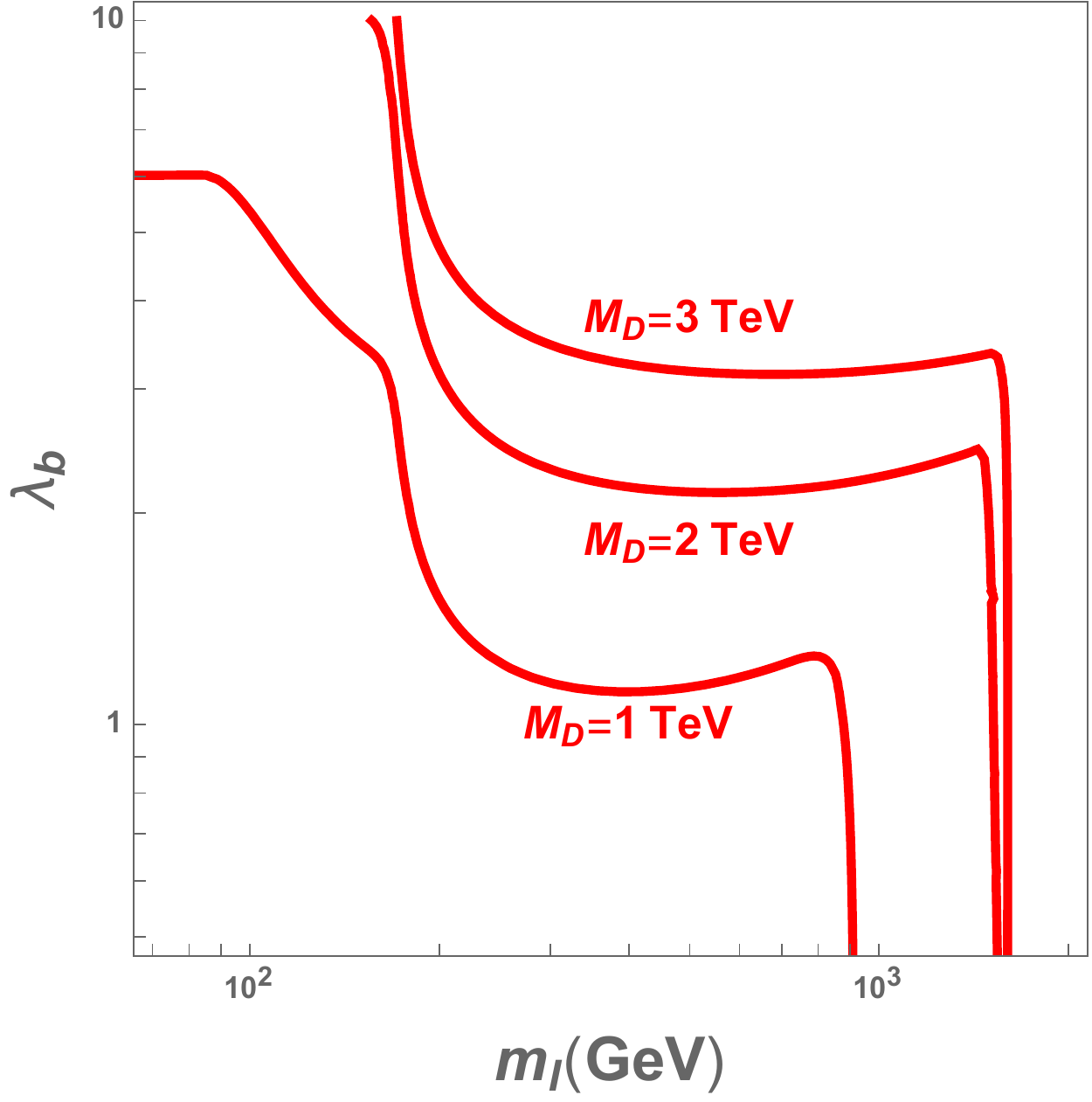}
\end{center}
\caption{A plot of $\lambda_b$ as a function of $m_I$ to give constant $\Omega_{\rm DM} h^2=0.1199$ for three different values of
  $M_D$: $M_D=1,  2, 3$ TeV. The rest of the fixed parameters can be found in the text.}
\label{fig:MI-lambdab}
\end{figure}

Now we consider the parameter space where type (b) diagrams in Fig.~\ref{fig:relic}  dominate the annihilation cross section.
Fig.~\ref{fig:MI-lambdab} shows the dependence of the dark Yukawa coupling $\lambda_b$ as a function of $m_I$ to give
constant $\Omega_{\rm DM} h^2=0.1199$ for three different values of $M_D$: $M_D=1, 2, 3$ TeV. 
For this we take heavy $m_{Z'}=10$ TeV and $m_{H_2}=10$ TeV to suppress the $X_I X_I \to Z' Z'$ channel. 
We set $\lambda_s=0.4$.
For the other fixed parameters, we take the same values with those for Fig.~\ref{fig:MI-alphaX}.
The $\lambda_b$ required to give $\Omega_{\rm DM} h^2$ changes sharply near $m_t/2$ and $m_t$.
This occurs due to the processes $X_I X_I \to c \bar{t} (\bar{c} t)$ and $X_I X_I \to t \bar{t}$, respectively.
Their $s$-wave annihilation cross sections are given by
\begin{align}
\sigma v (X_I X_I \to c \bar{t} + \bar{c} t) &= \frac{3 |\lambda_c|^2 |\lambda_t|^2 m_t^2 (4 m_I^2 -m_t)^2}{64 \pi 
m_I^4 (2 M_U^2+2 m_I^2-m_t^2)} + {\cal O}(v^2), \nl
\sigma v (X_I X_I \to t \bar{t} ) &= \frac{3 |\lambda_t|^4 m_t^2 (m_I^2 -m_t)^{3/2}}{16\pi m_I^3 (M_U^2+m_I^2-m_t^2)}+ {\cal O}(v^2),
\end{align}
where we have neglected the mass of charm quark. Note that both are proportional to $m_t^2$.
The contribution from $X_I X_I \to b \bar{b}$, being proportional to $m_b^2$, is negligible compared to the above two 
processes\footnote{In the limit $m_b \to 0$, the leading term in $v$ is proportional to $v^4$, {\it i.e.} $d$-wave.}.
When $m_I \approx M_D$, the near vertical lines are due to the coannihilation processes such as
$D \bar{D} (U \bar{U})\to g g, q \bar{q}, Z g$  $(q=u, d, s, c, b, t)$, $ U \bar{D} \to W^+ g$, and $X_I D \to s g$.
They are sensitive to the SM $SU(3)_C$ gauge coupling and independent of $\lambda_b$.
If we require $\lambda_b$ to be of order 1, to give the correct relic density $M_D$ should not be much heavier than  ${\cal O}(1)$ TeV, 
increasing the prospect of producing $D$ or $U$ at the LHC.

\section{Interplay between the $\bsmm$ anomaly and the dark matter}

Now we investigate whether the parameter space which solves the $\bsmm$ anomaly can also give the correct relic density
for the dark matter. At this stage we impose the low energy flavour constraints discussed in the previous sections. We also consider constraints from
the direct detection experiments of dark matter such as LUX~\cite{Akerib:2016vxi}, PANDA~\cite{Cui:2017nnn}, and
XENON1T~\cite{Aprile:2018dbl}.
\begin{figure}[t]
\begin{center}
\includegraphics[width=.3\textwidth]{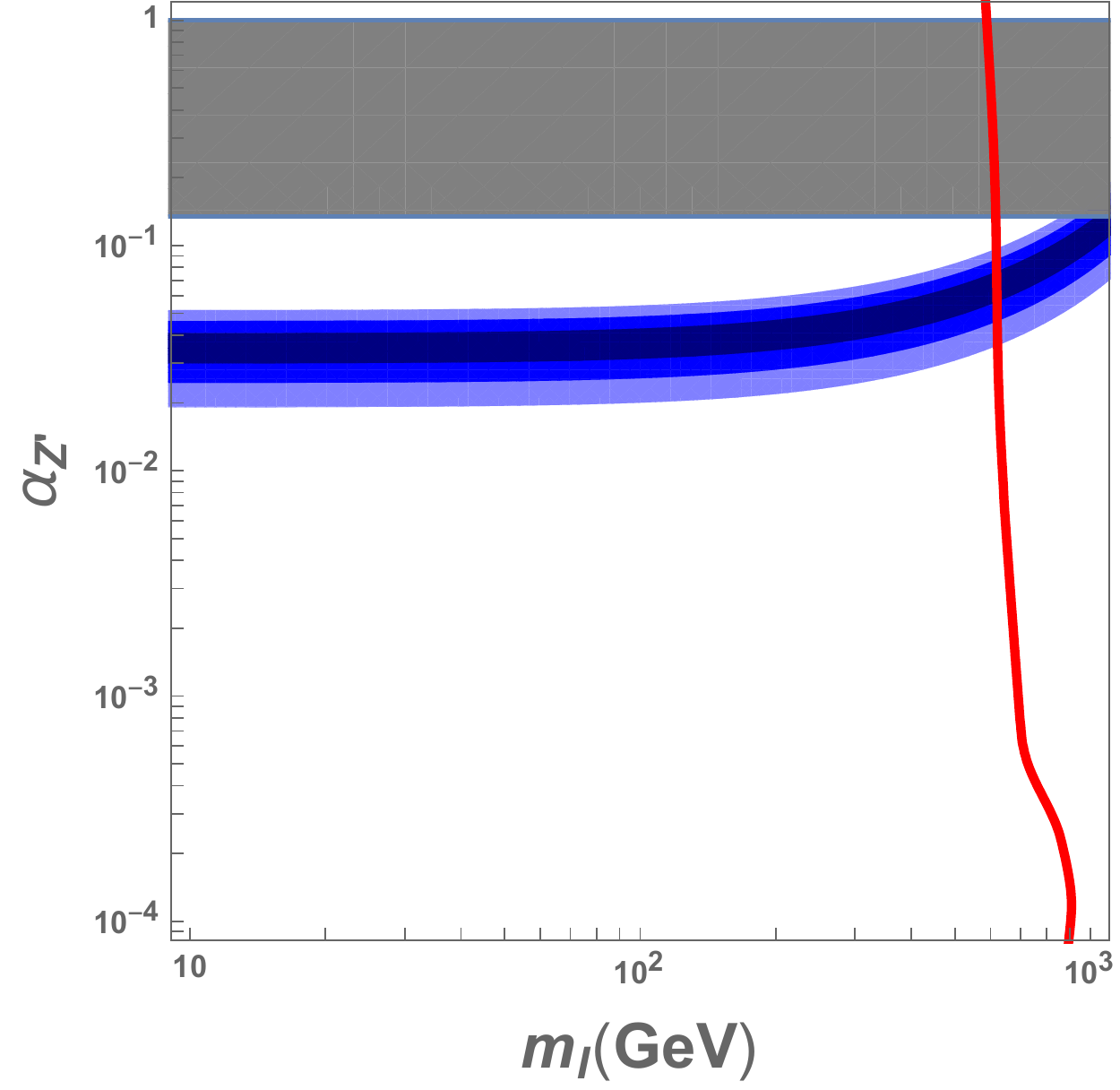}
\includegraphics[width=.3\textwidth]{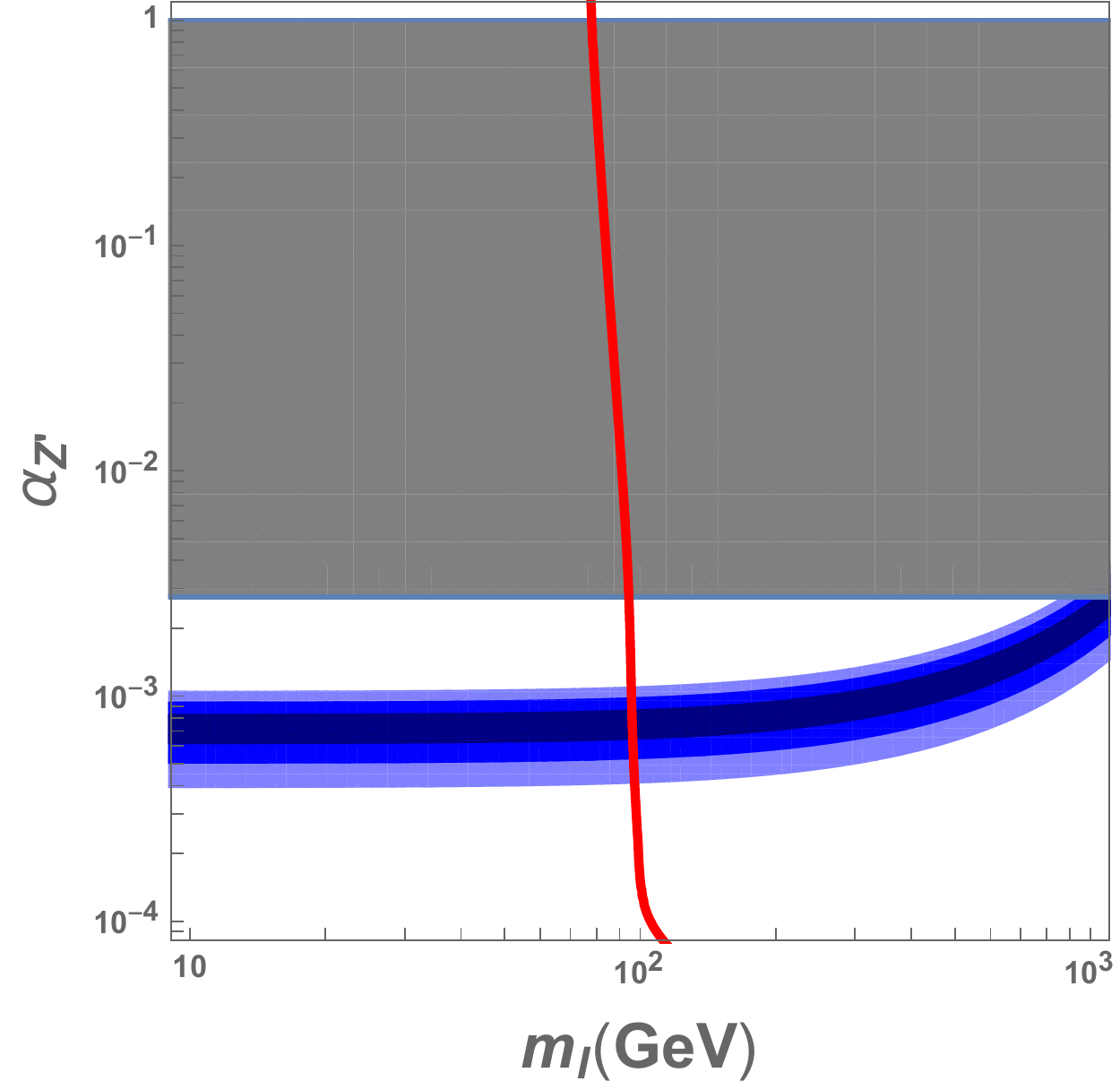}
\includegraphics[width=.3\textwidth]{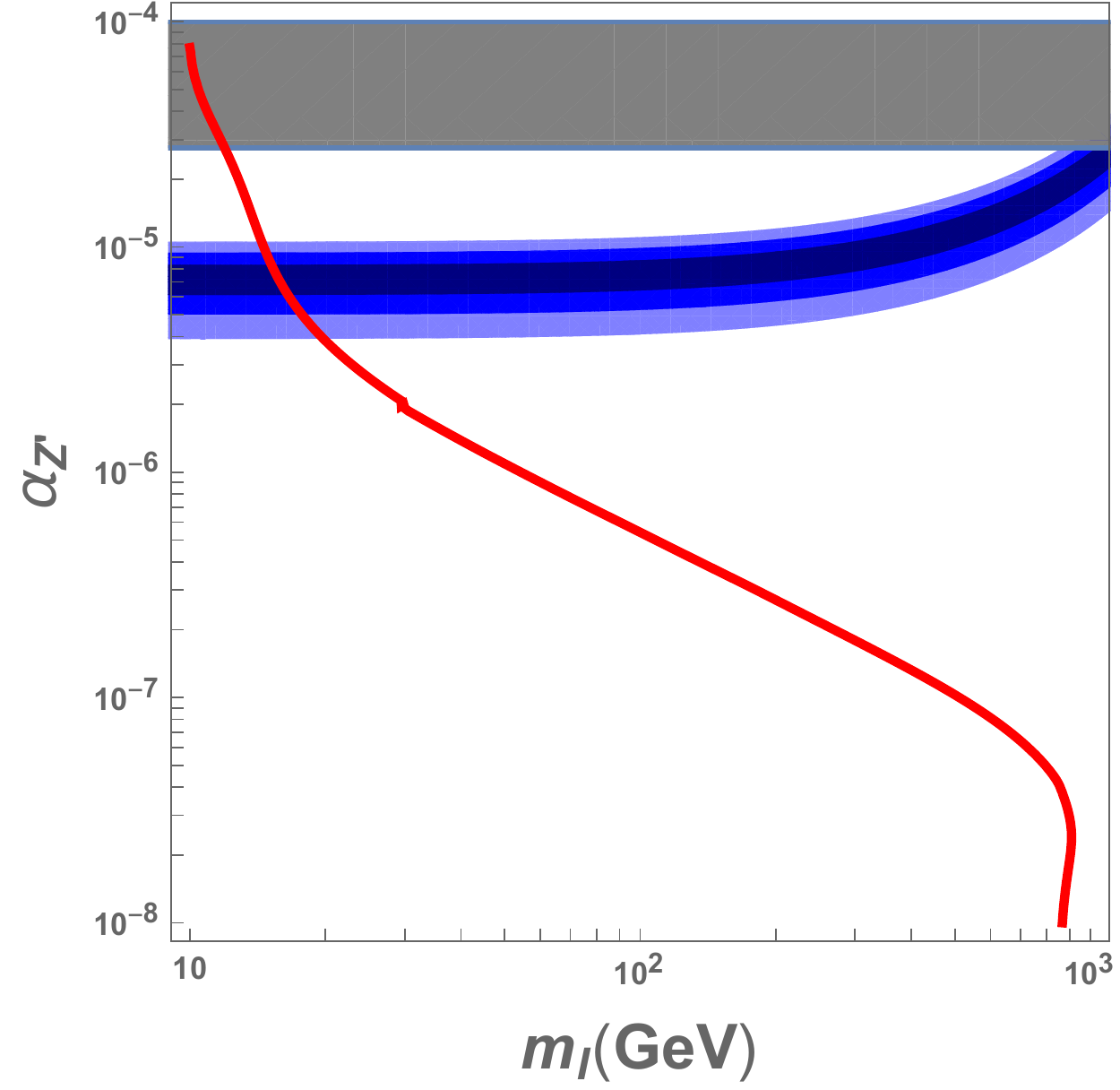}
\end{center}
\caption{The $C_9^{\mu,{\rm NP}}$ which solves the $\bsmm$ anomaly at 1$\sigma$ (dark blue), 
2$\sigma$ (blue), and 3$\sigma$ (light blue) in $(m_I,\alpha_{Z'})$ plane. We take $m_{Z'}=700, 100, 10$ GeV (from the left panel).
We fixed $q_X=2$, $\lambda_s \lambda_b^*=0.2$, $M_D=1$ TeV, and $m_R=3$ TeV.
They are superimposed with the constant  lines for $\Omega_{\rm DM} h^2=0.1199$ shown in Fig.~\ref{fig:MI-alphaX}. 
The grey regions are excluded by the neutrino trident production experiments at 2$\sigma$ level.}
\label{fig:MI-alphaX-C9}
\end{figure}

Fig.~\ref{fig:MI-alphaX-C9} shows plots for  $C_9^{\mu,{\rm NP}}$ which solves the $\bsmm$ anomaly at 1$\sigma$ (dark blue) and
2$\sigma$ (light blue) in $(m_I,\alpha_{Z'})$ plane. We take $m_{Z'}=700, 100, 10$ GeV (from the left panel).
We fixed $q_X=2$, $\lambda_s \lambda_b^*=0.2$, $M_D=1$ TeV, and $m_R=3$ TeV.
They are superimposed with the constant  lines for $\Omega_{\rm DM} h^2=0.1199$ shown in Fig.~\ref{fig:MI-alphaX}. 
The grey regions are excluded by the neutrino trident production experiments at 2$\sigma$ level.
We checked that the other low energy experiments do not further constrain the allows regions for the $C_9^{\mu,{\rm NP}}$ and the relic
density.
Neither does the direct detection experiments affect the plots in Fig.~\ref{fig:MI-alphaX-C9} because {\it i)} the $Z'$ couples to the quarks
at one-loop level and {\it ii)} more importantly  only inelastic upward scattering $X_I q \to X_R q$ can occur for $Z'$ interaction, which is
forbidden kinematically.
We can see that the $\bsmm$ anomaly can be resolved at 1$\sigma$ for $m_{Z'}=700$ GeV and the current dark matter
can be accommodated at the same time. For smaller $Z'$ masses, $m_{Z'}=100, 10$ GeV, $C_9^{\mu,{\rm NP}}$ becomes too large
and $\bsmm$ anomaly can be explained only at 2$\sigma$ level to explain the current relic density. 
This result shows a strong interplay between low energy $B$-meson decay experiments and the
dark matter physics.
\begin{figure}[t]
\begin{center}
\includegraphics[width=.3\textwidth]{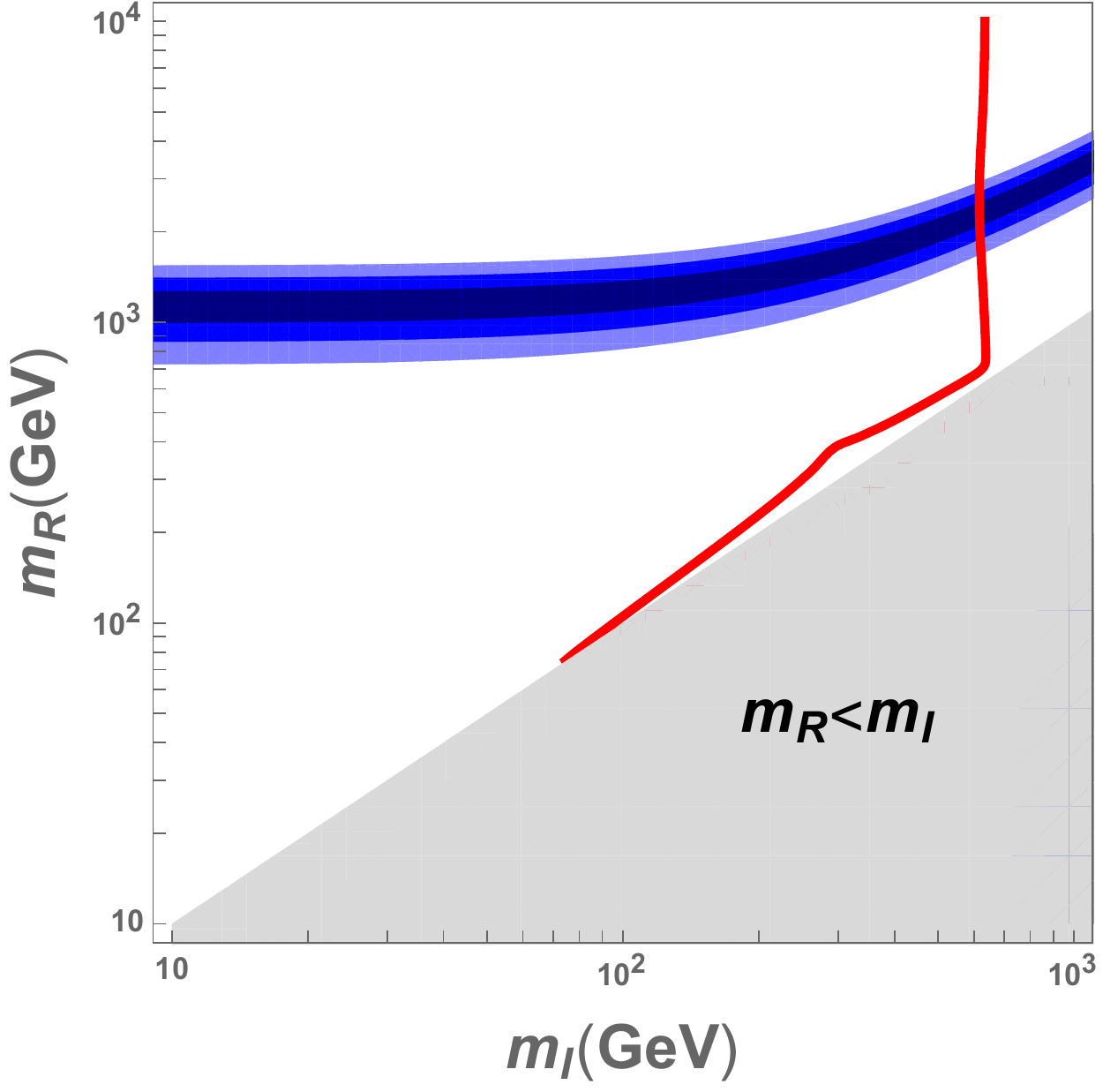}
\includegraphics[width=.3\textwidth]{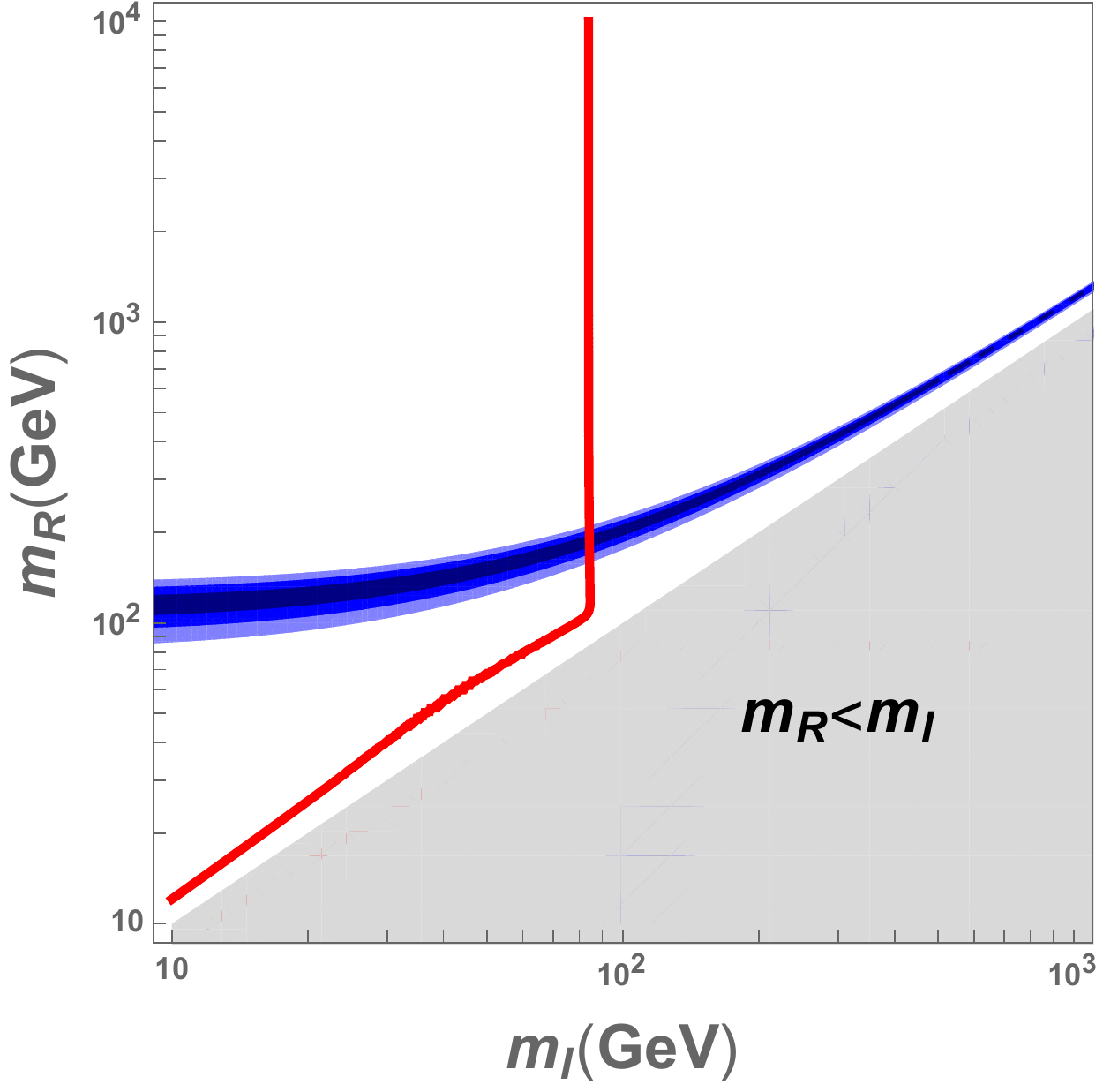}
\includegraphics[width=.3\textwidth]{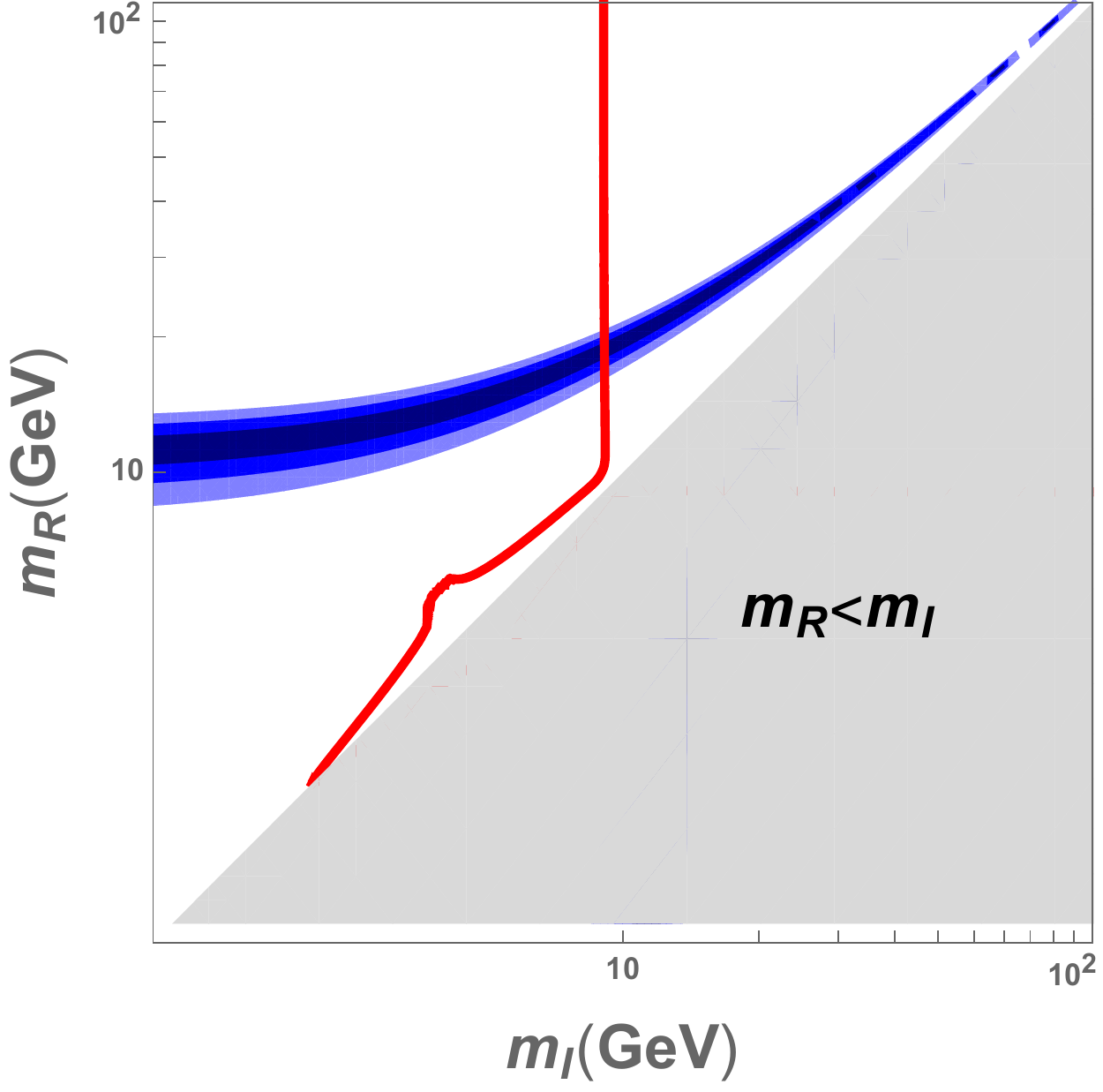}
\end{center}
\caption{The $C_9^{\mu,{\rm NP}}$ for $\bsmm$ anomaly (blue region) and the relic density of the dark matter
(red line) in $(m_I, m_R)$ plane. The dark blue, blue, and light blue region represent 1, 2, and 3$\sigma$ allowed region, respectively.
From the left panel we take $m_{Z'}=700, 100, 10$ GeV. We fixed
$q_X=2$, $\alpha_{Z'}=0.1$, $\lambda_s \lambda_b^*=0.2$, and $M_D=1$ TeV. The other fixed parameters are the same with those for 
Fig.~\ref{fig:MI-alphaX}. The grey region is unphysical because $m_R<m_I$.}
\label{fig:MI-MR-C9}
\end{figure}

In Fig.~\ref{fig:MI-MR-C9} we show the $C_9^{\mu,{\rm NP}}$ for $\bsmm$ anomaly (blue region) and the relic density of the dark matter
(red line) in
$(m_I, m_R)$ plane. From the left panel we take $m_{Z'}=700, 100, 10$ GeV. As in the previous case we fixed
$q_X=2$, $\alpha_{Z'}=0.1$, $\lambda_s \lambda_b^*=0.2$, and $M_D=1$ TeV. The other fixed parameters are the same with those for 
Fig.~\ref{fig:MI-alphaX}. The grey region is unphysical because $m_R <m_I$ and we exclude it.
The regions  we considered are not constrained by other observables such as $B_s-\bar{B}_s$ mixing, neutrino trident production,
or direct detection of dark matter.
For each case there is intersection region of the required $C_9^{\mu,{\rm NP}}$ and the correct relic density free from other
experimental constraints. The region occurs near the kinematic threshold of $X_I X_I \to Z' Z'$.
Relatively large $m_R$ compared to $m_I$ is also required  to get sizable $C_9^{\mu,{\rm NP}}$.

\begin{figure}[t]
\begin{center}
\includegraphics[width=.3\textwidth]{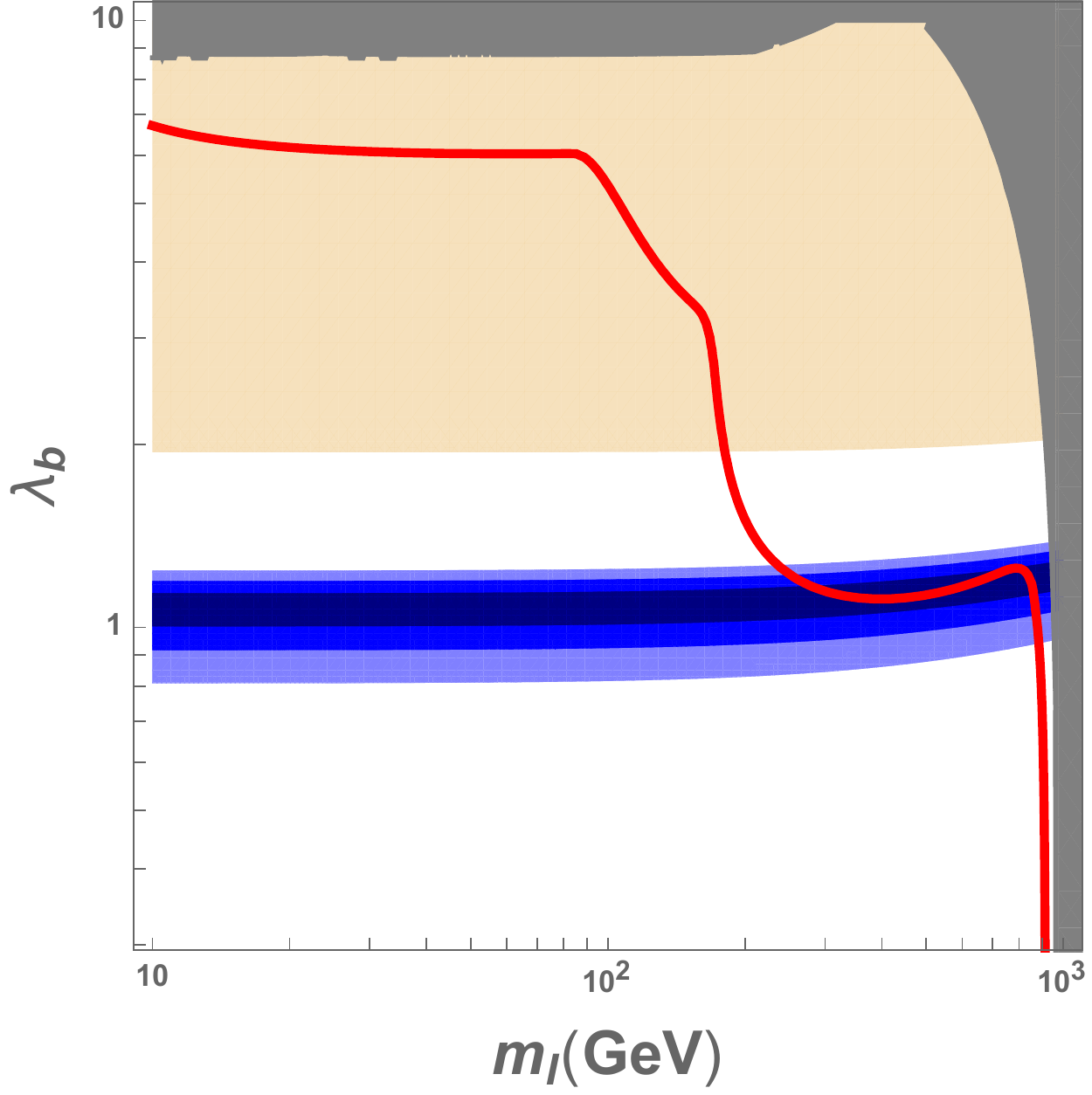}
\includegraphics[width=.3\textwidth]{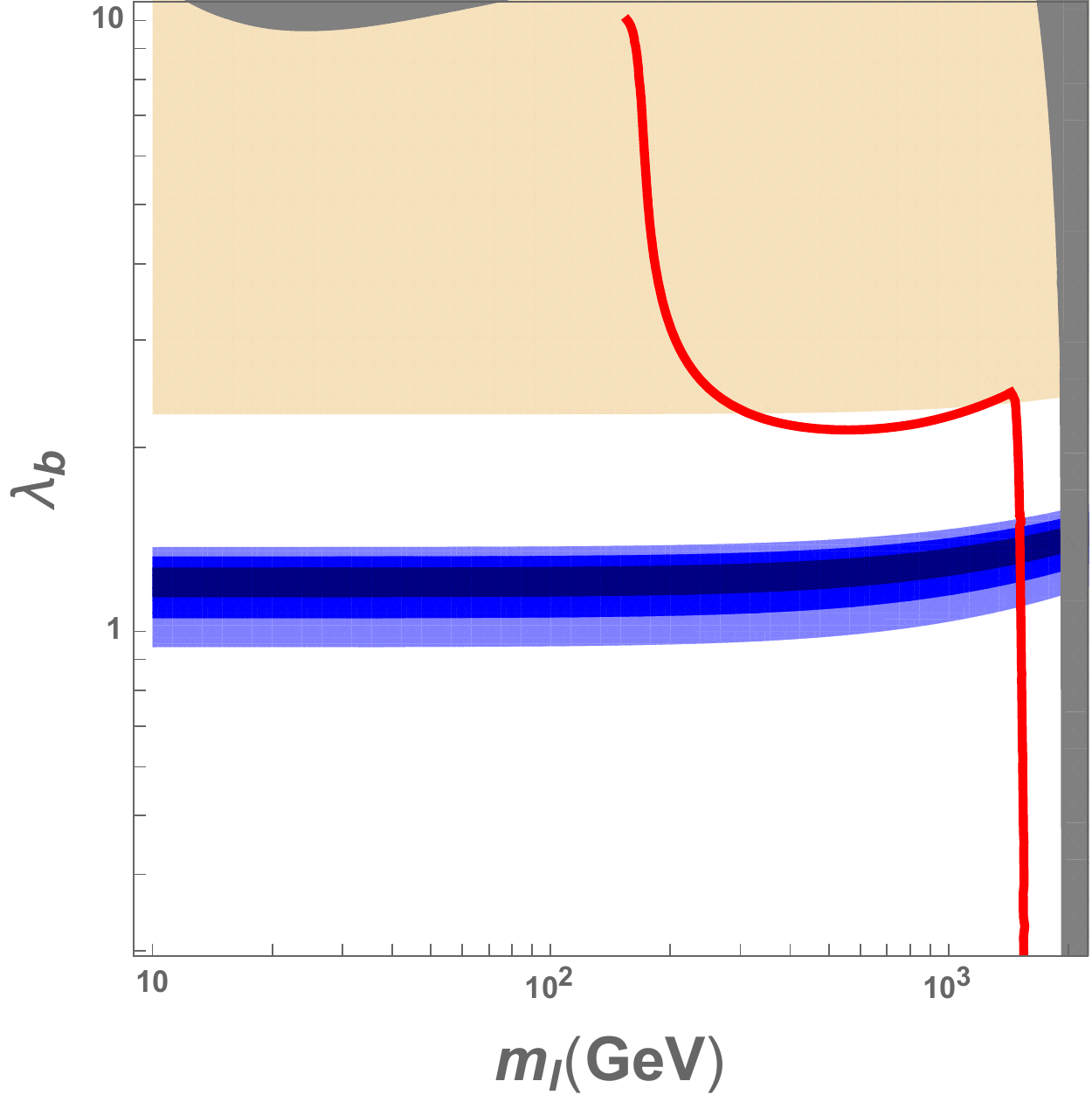}
\includegraphics[width=.3\textwidth]{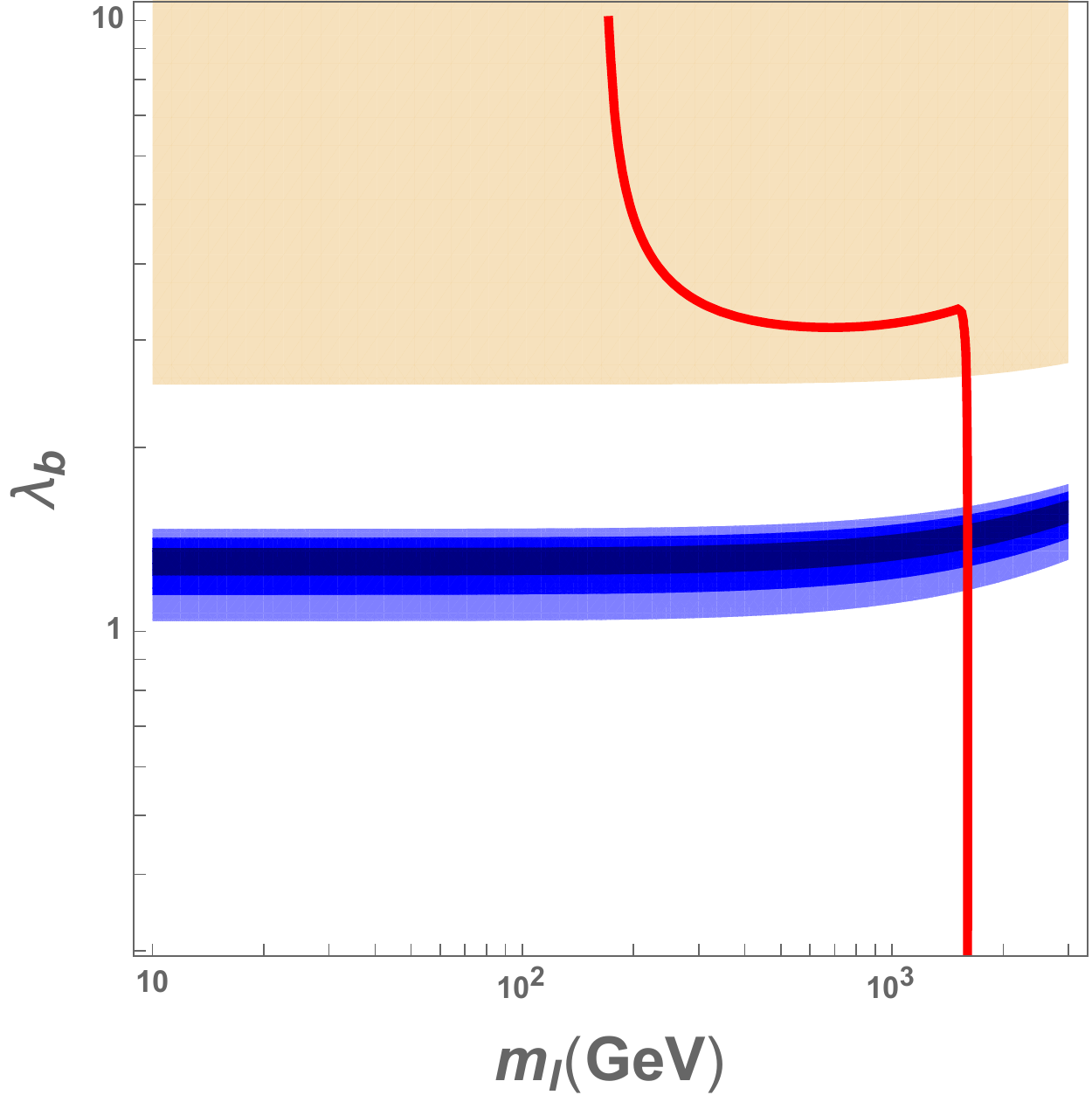}
\end{center}
\caption{The predictions of the DM relic density, $C_9^{\mu,{\rm NP}}$ in $(m_I,\lambda_b)$ plane. 
From the left panel we set $M_D=1, 1.5, 2$ TeV. We take $q_X=2$, $\alpha_{Z'}=0.1$, $m_{Z'}=2$ TeV, and $\lambda_s=0.4$, $m_R=3$ TeV.
 }
\label{fig:MI-lambdab-C9}
\end{figure}

Now we consider the impact of $\lambda$ couplings on the dark matter and the $C_9^{\mu,{\rm NP}}$. For this purpose
we suppress the dark gauge contributions to them by decoupling $Z'$ as in Fig.~\ref{fig:MI-lambdab}. 
To decouple we assume $Z'$ is heavy: for our purpose it is enough to set $m_{Z'}=2$ TeV as in Fig.~\ref{fig:MI-lambdab}. 
Fig.~\ref{fig:MI-lambdab-C9} shows the results
of this setting in $(m_I,\lambda_b)$ plane. We fixed $q_X=2$, $\alpha_{Z'}=0.1$, 
$\lambda_s=0.4$, $m_R=20$ TeV and $M_D=1, 2, 3$ TeV (from the left panel). 
We take the same values with those for Fig.~\ref{fig:MI-lambdab} for the other fixed parameters.
The grey region is excluded because the cross section of the DM scattering off the nuclei is too large.
For $M_D=3$ TeV the direct detection constraint disappears completely from the region considered.
The region with peach color is excluded by the experimental constraints on $\Delta m_s$ of $B_s -\bar{B}_s$ system.
We can see that the direct detection experiments and the $B_s-\bar{B}_s$ mixing play a complementary role in excluding the
parameter region, although the latter plays more important role in our choice of parameters.
Both the relic density and the $\bsmm$ anomaly can be explained simultaneously for the electroweak scale DM.
We notice that the contribution to $C_9^{\mu,{\rm NP}}$ is not easily decoupled for very heavy $m_{Z'}$ and $m_R$ due to
large mass splitting $m_R-m_I$.  Eventually as $m_{Z'}$ and/or $m_R$ becomes even heavier, their impact on $C_9^{\mu,{\rm NP}}$ will
get smaller. To see this decoupling effect we need to resum the large logarithm of $\log(m_R/M_D)$ and also consider higher loop
effects, which is beyond the current analysis.

\section{Conclusions}
\label{sec:concl}
{ We considered a new physics model with the $U(1)_{L_\mu-L_\tau}$ symmetry which has both a dark matter candidate and new  flavour changing
neutral currents in the quark sector. This opens up a possibility that there may exist a strong interplay between the dark matter and the flavour physics.
In particular we showed that we could simultaneously explain the $\bsmm$ anomaly and the dark matter abundance in our universe.
The model has a scalar dark matter candidate $X$, a $SU(2)_L$-doublet colored fermion $Q=(U, D)^T$, and a dark Higgs $S$ whose VEV breaks the
dark $\U1mt$ gauge symmetry spontaneously. Since the field $Q$ is vector-like under the gauge group, the model is free of the gauge anomaly.
Their charges are assigned in such a way that after the dark Higgs $S$ gets a VEV, there still remains a remnant
discrete $Z_2$ symmetry  of $\U1mt$.  After  the $\U1mt$ is broken down to the $Z_2$, the complex field $X$ is split into the two
real scalar fields  $X_I$ and $X_R$. The $X_I$, being the lightest $Z_2$-odd particle, is the dark matter candidate in the model. 
The other $Z_2$-odd particles are the $X_R$, and the $Q$.}

{ We identified some benchmark points which can explain both the $\bsmm$ anomaly and the relic abundance of dark matter.
We checked that they avoid
the known constraints from the $B$-meson decays, the $K$-meson decays, the LEP experiments, 
the measurements of the anomalous magnetic moment of muon, and also
the experiments of the direct detection of dark matter. When the $X_I X_I \to Z' Z'$ annihilation diagrams dominate, 
the $Z'$, $X_R$, $X_I$, $Q$ with the electroweak scale masses can explain both the DM relic density
and the $\bsmm$ anomaly while satisfying the current constraints. Since the $C_9^{\mu,{\rm NP}}$ for $\bsmm$ anomaly requires a sizable mass difference $m_R-m_I$, 
the coannihilation region $m_R \approx m_I$ is ruled out in this case.
When the $X_I X_I \to q \bar{q}$ mode dominates the relic density calculation, the $Z'$, $X_R$, and the dark Higgs can be much heavier than
the electroweak scale ($\gtrsim 10$ TeV) while the DM, $Q$ are still at the TeV scale.  When $\lambda_b \sim {\cal O}(1)$ the 
$C_9^{\mu, {\rm NP}}$ is still sizable because it is not easily decoupled with large $m_R-m_I$.
The DM relic density can be explained by either the $X_I X_I \to t \bar{t}$ channel or the $Q \bar{Q} \to {\rm SM} \; {\rm SM}$ channel.}

The model is a spin-flipped version of the model considered in \cite{Baek:2017sew} and shares some results in common.
In both models the $Z'$-penguin diagram can accommodate the required $C_9^{\mu,{\rm NP}}$ to explain the $\bsmm$ anomaly
and the dark matter candidate can explain the current relic density of the universe.
{The strongest flavour constraint comes from the mass difference in the $B_s-\bar{B}_s$ system in both models.
Compared with \cite{Baek:2017sew}, we included some new constraints such as the photon ($Z$-boson) penguin diagrams, the new physics
contributions to the $Z b \bar{b}$ vertex,
and extended the discussion on the dark matter.
}

\appendix
\section{Loop functions}
\label{app:loop}
The loop function $k(x)$ is defined as
\begin{align}
k(x) &= \frac{ x^2 \log x}{x-1}. 
\end{align}
We obtain $k(1)=1$.
When the number of arguments is greater than one, the loop function is defined recursively by
\begin{align}
k(x_1,x_2, x_3, \cdots) \equiv \frac{k(x_1,x_3, \cdots) - k(x_2, x_3, \cdots)}{x_1 - x_2}. 
\end{align}
From this definition, for example, we get
\begin{align}
k(x,x)=\frac{x (x-1+(x-2) \log x)}{(x-1)^2},   \quad
k(1,x)=\frac{-x+1+x^2 \log x}{(x-1)^2}, 
\end{align}
with $k(1,1)=3/2$.
The loop function $J_1(x)$ for $b \to s \gamma$ is 
\begin{align}
J_1(x) &=\frac{1-6x+3x^2+2x^3-6x^2 \log x}{12(1-x)^4}.
\end{align}
We have $J_1(1)=1/24$. The loop function for the photon-penguin of $b \to s \ell \bar{\ell}$ the effective $Z b \bar{b}$ vertex is
\begin{align}
Q_1(x) =\frac{7-36x+45 x^2-16 x^3-6 x^2(3-2x) \log x}{36(1-x)^4}.
\label{eq:loop_fn_Q1}
\end{align}
We get $Q_1(1)=1/8$.

\vspace{0.5cm}
\noindent
{\bf Note added:}
After finalizing the manuscript we received a paper considering $\bsmm$ anomaly in a similar but a different
setting~\cite{Hutauruk:2019crc}.
Their model does not have $\mu$ term, and as a consequence the dark matter candidate is complex scalar while it is
real scalar in our case. Their results show the allowed region is rather restricted compared to ours
due to the absence of the $\mu$ term.

\acknowledgments
This work was supported by the National Research Foundation of Korea(NRF) grant funded by the Korea government(MSIT) 
(Grant No. NRF-2018R1A2A3075605).

\bibliographystyle{JHEP}
\bibliography{RK_mt_SDM}

\end{document}